\documentclass[conference]{IEEEtran}

\ifCLASSINFOpdf
\else
\fi
\usepackage{tcolorbox}
\usepackage{tikz}
\usepackage{hyperref}
\usepackage{multirow}
\usepackage{listings}
\usepackage{xcolor}
\usepackage{amssymb}
\usepackage{amsmath}
\usepackage{cleveref}
\usepackage{balance}
\usepackage{makecell}

\Crefname{figure}{Fig.}{Figs.}

\usepackage{booktabs}       
\usepackage{colortbl}       

\definecolor{basebg}  {RGB}{247,251,247}   
\definecolor{flipbg}  {RGB}{255,247,247}   
\definecolor{rowgray} {RGB}{240,240,243}   

\lstdefinestyle{flip}{
    basicstyle=\ttfamily\scriptsize,
    breaklines=true,
    breakatwhitespace=false,
    columns=fullflexible,
    keepspaces=true,
    showstringspaces=false,
    aboveskip=0pt,
    belowskip=0pt,
    xleftmargin=4pt,
    xrightmargin=4pt,
    numbers=none,
}

\newcommand{\circnum}[1]{%
  \tikz[baseline=(char.base)]{
    \node[shape=circle,draw,fill=black,inner sep=1pt] (char)
    {\textcolor{white}{\bfseries #1}};}}

\newcommand{\rcircnum}[1]{%
  \tikz[baseline=(char.base)]{
    \node[shape=circle,draw=red,fill=red,inner sep=1pt] (char)
    {\textcolor{white}{\bfseries #1}};}}
\usepackage{array}
\usepackage{booktabs}
\usepackage{tabularx}
\usepackage[table]{xcolor}
\usepackage{graphicx}
\usepackage{xurl}

\usepackage{tabularx}
\usepackage{booktabs}
\usepackage{array}

\definecolor{IEEEBlue}{RGB}{31,78,121}

\newcounter{takeawayctr}
\definecolor{myorange}{rgb}{0.98, 0.69, 0.45}
\definecolor{mygreen}{rgb}{0.635, 0.933, 0.765}
\definecolor{myaliceblue}{rgb}{0.94, 0.97, 1.0}
\definecolor{myballblue}{rgb}{0.13, 0.67, 0.8}
\definecolor{mybrightmaroon}{rgb}{0.76, 0.13, 0.28}
\definecolor{mycadet}{rgb}{0.33, 0.41, 0.47}
\definecolor{mycadetblue}{rgb}{0.37, 0.62, 0.63}
\definecolor{mycharcoal}{rgb}{0.21, 0.27, 0.31}
\definecolor{mywarmblack}{rgb}{0.0, 0.26, 0.26}
\definecolor{myroyalblue}{rgb}{0.0, 0.14, 0.4}

\renewcommand{\thetakeawayctr}{\Roman{takeawayctr}}

\newcommand{\llvmhammer}{%
  \raisebox{-0.3ex}{\includegraphics[height=1.2em]{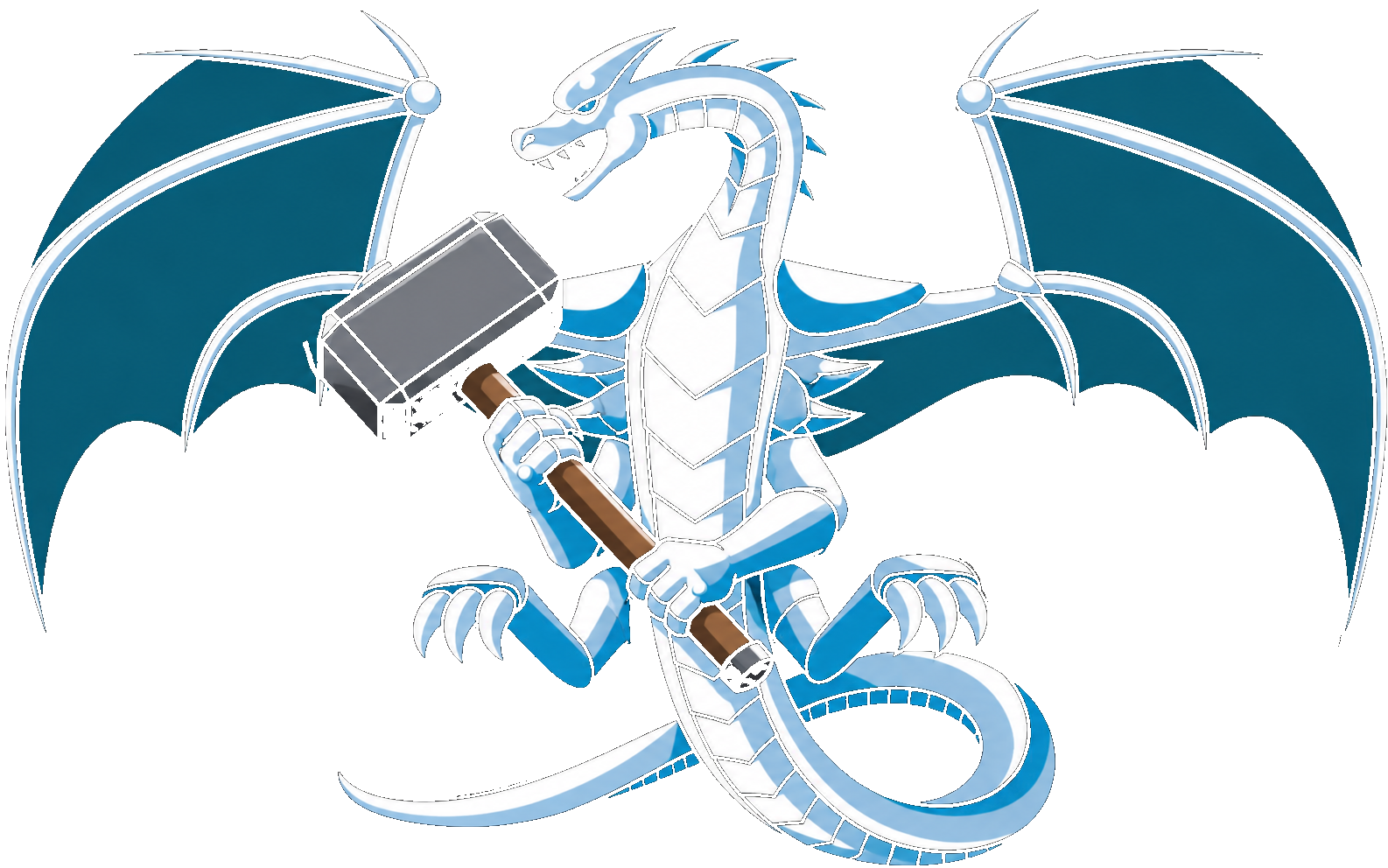}}%
}
\newcommand{\takeaway}[2]{%
\refstepcounter{takeawayctr}%
\vspace{5pt}
\noindent
\begin{tikzpicture}
\node[
    anchor=text,
    minimum width=\columnwidth-1pt,
    text width=\dimexpr\columnwidth - 14pt\relax,
    draw=myroyalblue,
    dotted,
    rounded corners=4pt,
    line width=1pt,
    fill=myaliceblue,
    inner xsep=5pt,
    inner ysep=10pt,
] (box) {#2};

\node[
    draw=myroyalblue,
    fill=myroyalblue,
    rounded corners=2pt,
    line width=1pt,
    anchor=west,
    inner xsep=3pt,
    inner ysep=2.5pt, 
    xshift=3mm,
    yshift=1mm
] at (box.north west) 
{\llvmhammer \textcolor{white}{\textbf{~Takeaway~\thetakeawayctr:~#1}}};

\end{tikzpicture}%
}

\newcommand{\emphBox}[2]{%
    \vspace{5pt}
    \noindent
    \begin{tikzpicture}

    \node[
        anchor=text,
        minimum width=\columnwidth-1pt,
        text width=\dimexpr\columnwidth - 14pt\relax,
        draw=myroyalblue,
        dotted,
        rounded corners=4pt,
        line width=1pt,
        fill=myaliceblue,
        inner xsep=5pt,
        inner ysep=10pt,
    ] (box) {#2};

    \node[
        draw=myroyalblue,
        fill=myroyalblue,
        rounded corners=2pt,
        line width=1pt,
        anchor=west,
        inner xsep=3pt,
        inner ysep=2.5pt, 
        xshift=3mm,
        yshift=1mm
    ] at (box.north west) 
    {\llvmhammer \textcolor{white}{\textbf{~#1}}};

    \end{tikzpicture}%
}

\begin{document}

\definecolor{codebg}      {RGB}{248,248,250}
\definecolor{flipbg}      {RGB}{255,247,247}
\definecolor{basebg}      {RGB}{247,251,247}
\definecolor{rowgray}     {RGB}{240,240,243}

\definecolor{llvm-kw}     {RGB}{ 86, 98,180}   
\definecolor{llvm-type}   {RGB}{160, 60,120}   
\definecolor{llvm-reg}    {RGB}{ 30,120, 80}   
\definecolor{llvm-str}    {RGB}{180, 80, 20}   
\definecolor{llvm-cmt}    {RGB}{140,140,150}   
\definecolor{llvm-num}    {RGB}{ 50, 90,160}   

\definecolor{ptx-op}      {RGB}{ 86, 98,180}   
\definecolor{ptx-mod}     {RGB}{160, 60,120}   
\definecolor{ptx-reg}     {RGB}{ 30,120, 80}   
\definecolor{ptx-pred}    {RGB}{180,100,  0}   
\definecolor{ptx-cmt}     {RGB}{140,140,150}   
\definecolor{ptx-num}     {RGB}{ 50, 90,160}   

\definecolor{sass-op}     {RGB}{ 86, 98,180}   
\definecolor{sass-mod}    {RGB}{160, 60,120}   
\definecolor{sass-reg}    {RGB}{ 30,120, 80}   
\definecolor{sass-pred}   {RGB}{180,100,  0}   
\definecolor{sass-addr}   {RGB}{ 40,140,140}   
\definecolor{sass-cmt}    {RGB}{140,140,150}   
\definecolor{sass-offset} {RGB}{ 50, 90,160}   

\lstdefinelanguage{LLVMIR}{
  sensitive=true,
  morecomment=[l]{;},
  morestring=[b]",
  morekeywords=[1]{tail,call,asm,sideeffect,ret,define,ptr,addrspace,
                   alloca,load,store,getelementptr,bitcast,zext,sext,
                   trunc,add,sub,mul,icmp,br,phi,select,unreachable,
                   nounwind,readnone,argmemonly,inaccessiblememonly},
  morekeywords=[2]{i1,i8,i16,i32,i64,i128,half,float,double,
                   void,label,metadata,token},
}

\lstdefinestyle{llvmir-base}{
  language=LLVMIR,
  basicstyle=\ttfamily\scriptsize,
  backgroundcolor=\color{basebg},
  commentstyle=\color{llvm-cmt}\itshape,
  stringstyle=\color{llvm-str},
  keywordstyle=[1]\color{llvm-kw}\bfseries,
  keywordstyle=[2]\color{llvm-type},
  literate=
    *{\%}{{{\color{llvm-reg}\%}}}1
     {0}{{\textcolor{llvm-num}{0}}}1
     {1}{{\textcolor{llvm-num}{1}}}1
     {2}{{\textcolor{llvm-num}{2}}}1,
  frame=single,
  framerule=0.4pt,
  rulecolor=\color{green!40!black!30},
  framesep=4pt,
  breaklines=true,
  columns=fullflexible,
  keepspaces=true,
  showstringspaces=false,
  aboveskip=0pt,
  belowskip=0pt,
  xleftmargin=4pt,
  xrightmargin=4pt,
  numbers=none,
}

\lstdefinestyle{llvmir-flip}{
  style=llvmir-base,
  backgroundcolor=\color{flipbg},
  rulecolor=\color{red!40!black!30},
}

\lstdefinelanguage{PTX}{
  sensitive=false,
  morecomment=[l]{//},
  morecomment=[l]{;},
  morekeywords=[1]{mov,ld,st,add,sub,mul,mad,fma,cvt,setp,
                   selp,bra,ret,call,bar,atom,red,tex,suld,sust,
                   wmma,mma,cp,async},
  morekeywords=[2]{u8,u16,u32,u64,s8,s16,s32,s64,
                   f16,f32,f64,b8,b16,b32,b64,
                   global,local,shared,param,const,generic,
                   v2,v4,wide,lo,hi},
}

\lstdefinestyle{ptx-base}{
  language=PTX,
  basicstyle=\ttfamily\scriptsize,
  backgroundcolor=\color{basebg},
  commentstyle=\color{ptx-cmt}\itshape,
  keywordstyle=[1]\color{ptx-op}\bfseries,
  keywordstyle=[2]\color{ptx-mod},
  literate=
    *{\%r}{{{\color{ptx-reg}\%r}}}2
     {\%rd}{{{\color{ptx-reg}\%rd}}}3
     {\%p}{{{\color{ptx-reg}\%p}}}2
     {@\%p}{{{\color{ptx-pred}@\%p}}}3
     {0x}{{{\color{ptx-num}0x}}}2,
  frame=single,
  framerule=0.4pt,
  rulecolor=\color{green!40!black!30},
  framesep=4pt,
  breaklines=true,
  columns=fullflexible,
  keepspaces=true,
  showstringspaces=false,
  aboveskip=0pt,
  belowskip=0pt,
  xleftmargin=4pt,
  xrightmargin=4pt,
  numbers=none,
}

\lstdefinestyle{ptx-flip}{
  style=ptx-base,
  backgroundcolor=\color{flipbg},
  rulecolor=\color{red!40!black!30},
}

\lstdefinelanguage{SASS}{
  sensitive=false,
  morecomment=[l]{;},
  morekeywords=[1]{LDG,STG,LDS,STS,IMAD,IADD,IMUL,FADD,FMUL,FFMA,
                   DADD,DMUL,DFMA,MOV,SEL,PRMT,SHFL,VOTE,BREV,
                   POPC,CLZ,FLO,IABS,INEG,USHF,SHF,BFE,BFI,
                   MUFU,RCP,SQRT,LG2,EX2,COS,SIN,
                   ISETP,FSETP,DSETP,BSETP,
                   BRA,RET,CALL,JMP,EXIT,
                   BAR,MEMBAR,ATOM,RED,
                   LEA,XMAD,HMMA,BMMA,DEPBAR,NOP,CS2R,S2R},
  morekeywords=[2]{E,128,U16,U32,S32,F32,F64,
                   WIDE,HI,LO,X,SY,GL,CTA,LS},
}

\lstdefinestyle{sass-base}{
  language=SASS,
  basicstyle=\ttfamily\scriptsize,
  backgroundcolor=\color{basebg},
  commentstyle=\color{sass-cmt}\itshape,
  keywordstyle=[1]\color{sass-op}\bfseries,
  keywordstyle=[2]\color{sass-mod},
  literate=
    *{R0}{{{\color{sass-reg}R0}}}2
     {R1}{{{\color{sass-reg}R1}}}2
     {R2}{{{\color{sass-reg}R2}}}2
     {R3}{{{\color{sass-reg}R3}}}2
     {R4}{{{\color{sass-reg}R4}}}2
     {R5}{{{\color{sass-reg}R5}}}2
     {R6}{{{\color{sass-reg}R6}}}2
     {R7}{{{\color{sass-reg}R7}}}2
     {R8}{{{\color{sass-reg}R8}}}2
     {R9}{{{\color{sass-reg}R9}}}2
     {RZ}{{{\color{sass-reg}RZ}}}2
     {URZ}{{{\color{sass-reg}URZ}}}3
     {@P}{{{\color{sass-pred}@P}}}2
     {@!P}{{{\color{sass-pred}@!P}}}3
     {[}{{{\color{sass-addr}[}}}1
     {]}{{{\color{sass-addr}]}}}1
     {0x}{{{\color{sass-offset}0x}}}2,
  frame=single,
  framerule=0.4pt,
  rulecolor=\color{green!40!black!30},
  framesep=4pt,
  breaklines=true,
  columns=fullflexible,
  keepspaces=true,
  showstringspaces=false,
  aboveskip=0pt,
  belowskip=0pt,
  xleftmargin=4pt,
  xrightmargin=4pt,
  numbers=none,
}

\lstdefinestyle{sass-flip}{
  style=sass-base,
  backgroundcolor=\color{flipbg},
  rulecolor=\color{red!40!black!30},
}

\newsavebox{\llvmbase}\newsavebox{\llvmflip}
\newsavebox{\ptxbase}\newsavebox{\ptxflip}
\newsavebox{\sassbase}\newsavebox{\sassflip}
\begin{lrbox}{\llvmbase}
\begin{minipage}{0.38\textwidth}
\begin{lstlisting}[style=llvmir-base]
; single vectorized load (4x32-bit)
%val = tail call { i32, i32, i32, i32 }
  asm sideeffect
  "ld.global.v4.b32
     { $0, $1, $2, $3 }, [ $4 ];",
  "=r,=r,=r,=r,l"
  (ptr addrspace(1) %ptr) #2
\end{lstlisting}
\vspace{2pt}
{\scriptsize\textit{One vectorized inline-asm call loads 4$\times$32-bit values.}}
\end{minipage}
\end{lrbox}

\begin{lrbox}{\llvmflip}
\begin{minipage}{0.38\textwidth}
\begin{lstlisting}[style=llvmir-flip]
; four separate scalar loads
%v0 = tail call i32 asm sideeffect
  "ld.global.b32 { $0 }, [ $1 ];",
  "=r,l"(ptr addrspace(1) %ptr0) #6
%v1 = tail call i32 asm sideeffect
  "ld.global.b32 { $0 }, [ $1 ];",
  "=r,l"(ptr addrspace(1) %ptr1) #6
; ... two more scalar calls
\end{lstlisting}
\vspace{2pt}
{\scriptsize\textit{A single four-element vector load is scalarized into four scalar loads.}}
\end{minipage}
\end{lrbox}

\begin{lrbox}{\ptxbase}
\begin{minipage}{0.38\textwidth}
\begin{lstlisting}[style=ptx-base]
; 128-bit vectorized load
@%p0 ld.global.v4.b32
    { %r0, %r1, %r2, %r3 },
    [ %rd0 + 0 ];

; 128-bit vectorized store
@%p1 st.global.v4.b32
    [ %rd1 + 0 ],
    { %r4, %r5, %r6, %r7 };
\end{lstlisting}
\vspace{2pt}
{\scriptsize\textit{128-bit coalesced memory transactions; four values per instruction.}}
\end{minipage}
\end{lrbox}

\begin{lrbox}{\ptxflip}
\begin{minipage}{0.38\textwidth}
\begin{lstlisting}[style=ptx-flip]
; four scalar 32-bit loads
@%p0 ld.global.b32 { %r0 }, [ %rd0 + 0 ];
@%p0 ld.global.b32 { %r1 }, [ %rd0 + 4 ];
@%p0 ld.global.b32 { %r2 }, [ %rd0 + 8 ];
@%p0 ld.global.b32 { %r3 }, [ %rd0 + 12 ];

; four scalar 32-bit stores
@%p1 st.global.b32 [ %rd1 + 0  ], %r4;
@%p1 st.global.b32 [ %rd1 + 4  ], %r5;
\end{lstlisting}
\vspace{2pt}
{\scriptsize\textit{One 32-bit scalar load per value, resulting in 4$\times$ more memory transactions.}}
\end{minipage}
\end{lrbox}

\begin{lrbox}{\sassbase}
\begin{minipage}{0.38\textwidth}
\begin{lstlisting}[style=sass-base]
; 128-bit vectorized global loads
@P0 LDG.E.128 R4,  [R2.64];
@P0 LDG.E.128 R8,  [R6.64];

; 128-bit vectorized global stores
@P1 STG.E.128 [R10.64], R4;
@P1 STG.E.128 [R12.64], R8;
\end{lstlisting}
\vspace{2pt}
{\scriptsize\textit{Each LDG/STG.E.128 loads/stores 128 bits in one transaction.}}
\end{minipage}
\end{lrbox}

\begin{lrbox}{\sassflip}
\begin{minipage}{0.38\textwidth}
\begin{lstlisting}[style=sass-flip]
; 16-bit scalar loads replace 128-bit loads
@P0 LDG.E.U16 R4,  [R2.64];
@P0 LDG.E.U16 R5,  [R2.64+0x10];
@P0 LDG.E.U16 R6,  [R2.64+0x20];
@P0 LDG.E.U16 R7,  [R2.64+0x30];
; ... 48 x LDG.E.U16 total
; ... 0 x LDG/STG.E.128 remain
\end{lstlisting}
\vspace{2pt}
{\scriptsize\textit{Vectorized loads are replaced with multiple scalar loads.}}
\end{minipage}
\end{lrbox}

%
\title{Amplifying Latency and Cost of LLM Inference via Rowhammer Attack}

\author{\IEEEauthorblockN{Mrityunjay Shukla}
	\IEEEauthorblockA{Indian Institute of Technology Bombay \\Mumbai, India\\
		mrityunjay@cse.iitb.ac.in}
	\and
	\IEEEauthorblockN{Shubham Roy}
	\IEEEauthorblockA{Indian Institute of Technology Bombay \\Mumbai, India\\
		royshubham@cse.iitb.ac.in}
	\and
	\IEEEauthorblockN{Biswabandan Panda}
	\IEEEauthorblockA{Indian Institute of Technology Bombay \\Mumbai, India\\
		biswa@cse.iitb.ac.in}}

\maketitle


\begin{abstract}
Large language models (LLMs) achieve high inference efficiency using optimized GPU kernels generated by just-in-time (JIT) compilation frameworks such as Triton. Triton compiles GPU kernels through an MLIR/LLVM-based pipeline. We investigate whether a Rowhammer-induced bit-flip in the CPU's DRAM can degrade GPU-based LLM inference performance while preserving output correctness.
We present LLVMhammer, the first Rowhammer attack that causes LLM inference slowdown by inducing a bit-flip in Triton's compilation pipeline. A bit-flip in a performance-critical optimization pass causes Triton to generate a functionally identical but suboptimal GPU kernel. Since Triton caches the generated kernel on disk, a single successful bit-flip causes persistent slowdown across subsequent inference executions. Triton's target-independent stages are shared across GPU vendors, which makes our attack GPU vendor-agnostic. To identify exploitable bit-flip locations without manual inspection, we build Karnage, a tool that isolates useful bit-flip locations in Triton's target-independent compilation pipeline. We evaluate LLVMHammer across a diverse set of LLMs, including Qwen2.5-7B-Instruct, Llama-3.1-8B-Instruct, Mistral-7B-Instruct-v0.3, and Deepseek-llm-7B-chat, across NVIDIA RTX 4090 and AMD RX 7900 XTX GPUs. LLVMhammer increases inference latency of an LLM inference server running on a single GPU by 4-16\%, which, in turn, can translate into additional annual operating costs at the scale of multi-GPU deployments. We further show that LLVMhammer can also corrupt LLM outputs across different GPU vendors, reducing BERTScore to as low as 25\%, illustrating the broader security implications of bit-flips in Triton's compilation pipeline.

\end{abstract}


%
\IEEEpeerreviewmaketitle

\section{Introduction}
Large language models (LLMs) have become ubiquitous across a wide
range of applications, from conversational AI and code generation~\cite{llm_popularity_nvidia} to document summarization~\cite{llm_document_summarization}. The demand for LLM-powered
services continues to grow rapidly~\cite{llm_powered_services}. The effectiveness of an LLM deployment
depends on both output correctness and inference efficiency.
Correctness determines whether the model produces accurate and
reliable responses. Inference efficiency is determined by how effectively the deployment utilizes its computational resources and consequently, the throughput it can achieve, measured in terms of the number of tokens generated per second and the number of requests served per second. Lower inference efficiency reduces throughput and increases
operational costs. Modern LLMs achieve high inference efficiency by
executing optimized GPU kernels for matrix multiplication and
attention computations. The GPU software stack generates and manages
these kernels and therefore plays a critical role in every production
LLM serving system.


Rowhammer~\cite{flippingbits} is one of the attacks that can affect output correctness. A large body of research investigates accuracy degradation attacks on ML models~\cite{tbfa,bfa,terminal_brain_damage,yes_one_flip,deephammer,compiled_models_dnn-ndss26,sbfa,genbfa,bithydra-inference-token,flip_llm_host-2026,dontknock,gpurowhammer,prowhammer}. Most attacks target model weights~\cite{tbfa,bfa,terminal_brain_damage,deephammer,flip_llm_host-2026,bithydra-inference-token,genbfa,dontknock} and therefore require multiple bit-flips. Other works demonstrate accuracy degradation attacks under more practical and realistic assumptions using Rowhammer~\cite{yes_one_flip,deephammer,terminal_brain_damage,gpurowhammer,prowhammer}.
Some of the recent attacks~\cite{prowhammer, gpubreach} extend this threat to LLM inference by corrupting GPU
code. These attacks primarily target output integrity, causing incorrect predictions or incoherent responses. In contrast, we target inference efficiency, degrading LLM serving performance and thereby increasing deployment costs. 

\emphBox{Pertinent question}{Is it possible for a Rowhammer-induced bit-flip to reduce the inference efficiency of an LLM deployment running on GPUs without affecting the output correctness?}



\textbf {Our approach.} We answer the pertinent question by targeting vLLM~\cite{vLLM} (a widely used GPU \emph{vendor-agnostic} framework for production LLM inference) that relies on OpenAI's Triton~\cite{triton2019} (a Python-based domain-specific language and
compiler) to generate optimized GPU kernels for attention and matrix multiplication. Although NVIDIA's TensorRT-LLM~\cite{nvidia_tensorRTLLM} also relies on Triton, it only supports NVIDIA GPUs~\cite{tensorRTLLM_vs_vLLM}, thus reducing its applicability across different GPU vendors. Triton compiles these kernels \emph{just-in-time} (JIT) via the Multi-Level Intermediate Representation (MLIR) and LLVM compilation pipeline~\cite{mlir,llvm}. Triton stores the compiled kernels in a software cache located at \texttt{\textasciitilde/.triton/cache}~\cite{triton_github} and reuses them across future inference requests without recompilation.

We show that a Rowhammer-induced bit-flip in the Triton binary while it is resident in the CPU's DRAM can alter the generated GPU kernel, resulting in a suboptimal kernel while preserving output correctness. Triton then caches and reuses this suboptimal kernel, causing LLM inference slowdown. Overall, a single bit-flip persistently reduces inference efficiency in the form of a degradation in Time-to-first-token (TTFT)~\cite{vllm_benchmark}, Time-per-output-token (TPOT)~\cite{vllm_benchmark}, Inter-token-latency (ITL)~\cite{vllm_benchmark}, and throughput~\cite{ttft_tpot_itl_throughput}.

\textbf{Triton's compilation pipeline as an attack target.} Triton's JIT compilation pipeline is invoked during LLM inference, while its on-disk kernel cache allows a single corrupted compilation to \emph{persist} across subsequent requests and process restarts. Triton’s MLIR/LLVM-based compilation pipeline includes \emph{target-independent stages} shared across GPU vendors. The combination of JIT compilation, persistent kernel caching, and shared target-independent stages makes Triton's compilation pipeline an attack target. To the best of our knowledge, no prior work has examined Triton’s compilation pipeline as an attack target. 


\emphBox{Unexplored attack target}{
Triton’s target-independent optimization stages expose a previously unexplored attack target. As these optimizations are shared across GPU vendors, a single bit-flip can alter the generated kernels across multiple GPU vendors, causing LLM inference slowdowns.
}


\noindent\textbf{Challenges.} 
To mount such an attack on Triton, we have to overcome two major challenges. 

\noindent(i) \textit{Identifying target-independent functions}: The Triton library contains
more than 120,000 functions. Only a small subset of those functions implements the target-independent optimization passes that execute identically across all GPU vendors and kernel types.
Manually inspecting and isolating target-independent functions is intractable at this scale. 

\noindent(ii) \textit{Finding useful bit-flip locations within target-independent functions}: even after isolating the target-independent functions, each function contains many potential bit-flip locations, yielding
approximately 1.1-1.3 million potential bit-flip locations in total. Evaluating each potential bit-flip location requires inducing a bit-flip, compiling the kernel, and running a complete LLM inference, which takes 30 seconds for one location, making an exhaustive evaluation time-consuming (409 days). Furthermore, the vast
majority of potential bit-flip locations either leave the compiled kernel unchanged, crash it, or corrupt its output. Only a rare subset of bit-flip locations produce a kernel that is slower than the optimized baseline, correct in its outputs, and consistent in its slowdown across varying prompts and models. To address these challenges, we develop a tool called Karnage (\Cref{subsec:tool}). Karnage efficiently isolates useful bit-flip locations in Triton’s target-independent pipeline without manual inspection of bit-flip locations.

\begin{figure}[t]
    \centering
    \includegraphics[scale=0.8]{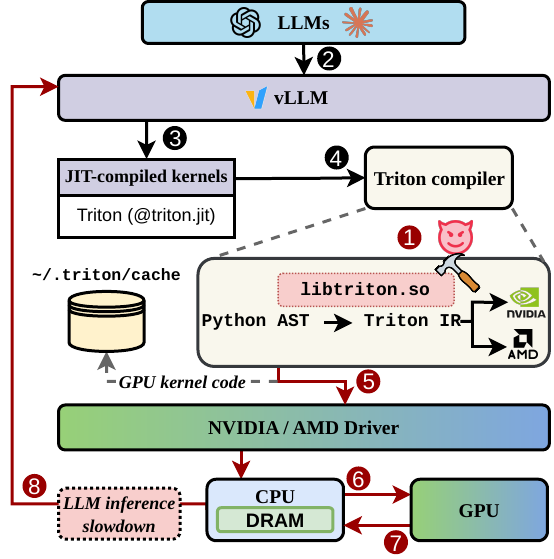}
    \caption{LLVMhammer in a nutshell.}
    \label{fig:attack_flow}
\end{figure}

\textbf{LLVMhammer.}~\Cref{fig:attack_flow} provides an overview of LLVMhammer, our proposed attack on LLM serving systems. The attacker's objective is to reduce
the inference efficiency of the LLM application in the form of a degradation in TTFT, TPOT, ITL, and throughput (tokens/second and requests/second). Before the LLM inference begins, in step (\rcircnum{1}), the attacker, a CPU process, uses Rowhammer to induce a bit-flip in a performance-critical optimization pass within the Triton compiler library (\texttt{libtriton.so}) resident in the CPU's DRAM. This
bit-flip alters the compiler's optimization decision and generates a suboptimal GPU kernel that preserves functional correctness while
degrading execution time. Triton stores generated GPU kernels in a software cache called the Triton cache~\cite{triton_github} and reuses them across future inference requests without recompilation. Therefore, a single
successful bit-flip can persistently affect subsequent inference
executions that reuse the suboptimal GPU kernel, including executions after process restart. 

In step (\circnum{2}), the victim LLM application initiates inference through an LLM serving framework such as vLLM. In step (\circnum{3}), during the execution, vLLM invokes JIT-compiled GPU kernels generated by the Triton compiler~(\circnum{4}). In step~(\rcircnum{5}), the generated suboptimal GPU kernel is passed to the
appropriate GPU driver. In step (\rcircnum{6}), the runtime and driver load the kernel into GPU
memory for execution. In step~(\rcircnum{7}), the
GPU executes the suboptimal kernel and produces inference outputs.
Finally, in step~(\rcircnum{8}), the LLM framework receives the outputs
and observes a slowdown in LLM inference.

\textbf{Key results and impact.}
We demonstrate the LLVMhammer attack using GPUs from two different vendors, namely NVIDIA RTX 4090 and AMD RX 7900 XTX, on four LLM models (Qwen2.5-7B-Instruct~\cite{qwen}, Llama-3.1-8B-Instruct~\cite{llama}, Mistral-7B-Instruct-v0.3~\cite{mistral}, and
Deepseek-llm-7B-chat~\cite{deepseek}). We evaluate the efficacy of our attack using five metrics, i.e., TTFT, TPOT, ITL, and throughput (tokens/second and requests/second). Overall, our attack targets an LLM inference server and increases total inference time up to 16\% when running Deepseek-llm-7B-chat for a large number of requests (1,000 requests~\cite{vllm_number_of_prompts}), compared with the attack-free baseline. This result demonstrates that our attack can cause a measurable slowdown in end-to-end LLM serving performance. Production LLM inference servers handle millions of requests per day~\cite{production_llm_load_source,vllm_production_llm_load_source}. A modest 10\% increase in inference latency loses roughly 876 hours (37 days) of serving capacity per year without any output anomaly to prompt investigation. In production deployments, this loss translates into a direct, sustained monetary cost as long as the suboptimal GPU kernel remains in Triton's on-disk cache. Production deployments range from single-GPU replicas to multi-node clusters, with NVIDIA reference architectures defining scales of 8, 128, and 256 GPUs~\cite{hf-vllm,nvidia-h100-nvl-ra,nvidia-hgx-ra}. At present, H100 pricing is \$4.50-\$10.00/GPU-hour~\cite{hf-pricing}, which translates to \$3.9K-\$8.76K per GPU annually, scaling to \$1M-\$2.24M per year for a 256-GPU deployment.

\noindent\textbf{Contributions.}
\begin{itemize}

\item We demonstrate LLVMhammer, the first Rowhammer attack that causes LLM inference slowdown by inducing a bit-flip in Triton's compilation pipeline, causing the compiler to generate a suboptimal GPU kernel. The slowdown persists across inference sessions and process restarts via Triton's on-disk kernel cache (\Cref{sec:attack_overview}).

\item We build Karnage, a tool that identifies exploitable bit-flip locations in the target-independent stages of the Triton pipeline without manual inspection (\Cref{subsec:tool}).

\item We show that LLVMhammer is GPU-vendor agnostic: bit-flip locations in Triton's target-independent stages generalize across NVIDIA and AMD GPUs, causing a 4--16\% increase in inference latency across four LLM models (\Cref{sec:eval}).

\item We also show that LLVMhammer can corrupt LLM outputs across GPU vendors, reducing BERTScore~\cite{bertscore} to as low as 25\%, demonstrating that bit-flips in Triton's compilation pipeline affect output correctness (\Cref{sec:generality}).

\end{itemize}

\section{Background}\label{sec:background}

In this Section, we briefly cover the Rowhammer attack, followed by the necessary background on the MLIR/LLVM toolchain, which is pivotal in understanding the compilation pipeline of Triton. Finally, we discuss how state-of-the-art LLM inference frameworks, such as vLLM, use Triton to generate efficient kernels.

\begin{figure}[t!]
    \centering
    \includegraphics[width=\linewidth]{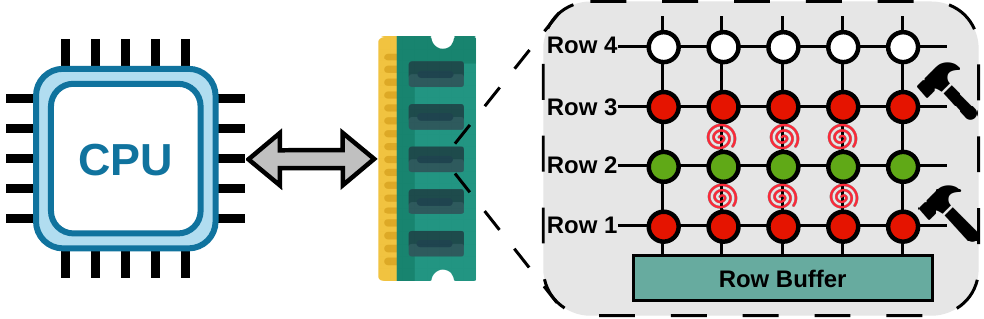}
    \caption{Rowhammer bug.}
    \label{fig:rowhammer}
\end{figure}

\begin{figure}[t]
    \centering
    \includegraphics[width=\linewidth]{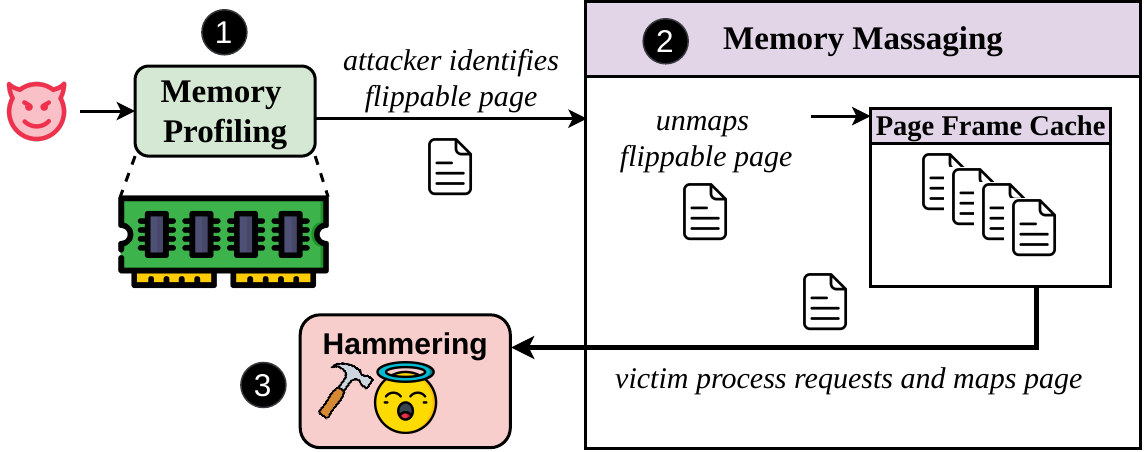}
    \caption{Steps in a Rowhammer attack.}
    \label{fig:attack_methodology}
\end{figure}

\subsection{Rowhammer attack}\label{subsec:rh_bc}
Rowhammer~\cite{flippingbits} is a DRAM integrity issue first demonstrated in 2014.
Over the past decade, Rowhammer has evolved into a versatile attack vector that spans security and privacy domains~\cite{flipfengshui,rambleed,blacksmith}.
The Rowhammer attack exploits cell-to-cell interference in modern DRAMs. Repeatedly activating (\emph{hammering}) one or more rows increases charge leakage in adjacent rows through capacitive coupling and electromagnetic interference, as shown in \Cref{fig:rowhammer}. When a DRAM row is activated sufficiently many times within a single refresh interval (typically 64 ms in DDR3 and DDR4 DRAMs and 32ms in DDR5 DRAMs), the affected neighboring cells may lose sufficient charge that, upon subsequent access, the sense amplifier misreads their stored value, resulting in a bit-flip. 
Attackers have exploited this reliability issue by deliberately accessing certain DRAM rows (called aggressor rows) in their own address space at high frequency, causing flips in the DRAM rows used by a victim process (called victim rows). DRAM vendors have implemented numerous countermeasures, such as Targeted Row Refresh (TRR)~\cite{trrespass} and the Per Row Activation Counter (PRAC) in DDR5~\cite{prac_jedec}, which refresh a victim row upon detecting suspicious row activations. However, TRR has been bypassed on several occasions~\cite{trrespass}~\cite{blacksmith}, and Rowhammer remains a threat even for the most recent DDR5 DRAMs~\cite{zenhammer}~\cite{ddr5_ecc_rowhammer}.

\begin{figure*}[t!]
    \centering
    \includegraphics[width=\linewidth]{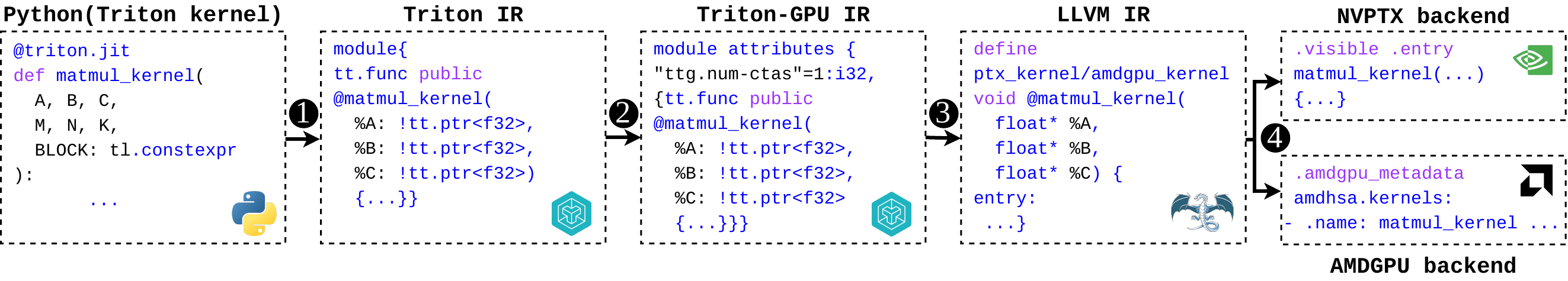}
    \caption{Triton compilation pipeline.}
    \label{fig:triton-pipeline}
\end{figure*}
 
To mount a successful Rowhammer attack, an attacker must follow three key steps~\cite{flipfengshui}, as shown in \Cref{fig:attack_methodology}.
The attack begins with \textbf{memory profiling} (\circnum{1}), in which the attacker profiles the DRAM to identify DRAM locations susceptible to bit-flips by allocating memory chunks from their own address space. It identifies the \emph{page-frames} (fixed-sized blocks in DRAM that hold an OS page) that have bit-flips at offsets suitable for their purpose.
Upon finding suitable bit-flip locations (i.e., at desired offsets within a DRAM page frame), the attacker manipulates the victim process's memory allocation so that the sensitive victim data or code is mapped to these locations. Typically, the OS's page-frame allocation policy, which allocates page frames without respecting process boundaries, is exploited in this step. This process is known as \textbf{memory massaging}  (\circnum{2}) ~\cite{sledgehammer,pqhammer,rambleed,yes_one_flip,prowhammer}. 
Finally, during the \textbf{hammering} phase (\circnum{3}), the attacker repeatedly accesses the identified aggressor rows in its own address space to induce bit-flips in the victim's rows. The end-to-end Rowhammer attack does not require any elevated privileges or root-level access to the target system.

\subsection{MLIR/LLVM intermediate representations}
A compiler intermediate representation (IR) provides a machine-independent program representation.
It lies between the source code and the final machine code.
The compiler frontend produces the IR after parsing the program.
The compiler backend converts the IR into machine instructions.
This separation simplifies compiler design and optimization.
The same optimizations work for many source languages
and work across different target architectures.

LLVM~\cite{llvm} provides a widely used intermediate representation called LLVM IR.
LLVM optimizes programs using a sequence of optimization passes.
Each pass analyzes or transforms the LLVM IR.
These passes reduce register pressure or instruction count.
Most LLVM passes are \emph{target-independent}.
LLVM passes do not depend on the target instruction set.
The same optimized IR supports many target instruction sets.
Examples include x86, ARM, AMDGPU, NVPTX, and RISC-V.
Only the final code generation depends on the target.

MLIR~\cite{mlir} extends this idea with multiple intermediate representations.
These representations are called \emph{dialects}.
Each dialect defines its own operations and types.
Multiple dialects coexist within a single compilation pipeline.
The compiler gradually compiles programs across these dialects.
Subsequent stages produce LLVM IR or target-specific representations.
MLIR optimization passes work on specific dialects.
Each pass only understands its own dialect and ignores lower target-specific dialects.
Therefore, the same optimizations work across different target backends.

\subsection{Triton and its compilation pipeline}

Triton~\cite{triton2019} is an open-source compiler and
Python-based domain-specific language (DSL) that lets developers
write GPU kernels without expertise in low-level GPU programming
models such as Compute Unified Device Architecture (CUDA) or Heterogeneous-computing Interface for Portability (HIP). Triton's compiler is responsible for architecture-specific optimizations such as register allocation, shared-memory placement, vectorization, and instruction selection. This
design has driven Triton's rapid adoption as the kernel compiler
for attention and matrix-multiplication kernels in LLMs' inference~\cite{vLLM}.

Triton's compilation pipeline compiles a kernel from the Python DSL to
target-specific machine code through four successive representations,
which \Cref{fig:triton-pipeline} illustrates. In step (\circnum{1}), the
compiler first converts the Python kernel into Triton IR, an
MLIR dialect that captures the blocked programming model directly,
operations in this dialect express computations over tiles, reductions
over blocks, and loads and stores from global and shared memory at a
level entirely independent of any target architecture.
In step (\circnum{2}), the compiler then compiles Triton IR into the
\textit{Triton-GPU dialect}, which introduces GPU-specific
abstractions such as thread layout annotations and shared memory
synchronization, but remains vendor-agnostic, i.e., it does not yet
distinguish between an NVIDIA warp and an AMD wavefront. The optimization passes at this level operate on GPU-specific abstractions while remaining largely independent of the target ISA, enabling the same optimization framework to support multiple GPU backends.
Next, in step~\circnum{3}, the compiler translates the Triton-GPU dialect into LLVM IR, at which point the representation specializes: for NVIDIA targets, it emits LLVM IR targeting the NVPTX backend; for AMD targets, it emits LLVM IR targeting the AMDGPU backend, with target-specific type annotations and intrinsics. In step~\circnum{4}, the LLVM backend emits target-specific machine code: NVIDIA's Parallel Thread Execution (PTX) for NVIDIA GPUs and AMD’s Graphics Core Next (AMDGCN) for AMD GPUs.

The boundary between (\circnum{2}) and (\circnum{3}) is the critical
architectural divide for this work. Every optimization pass that
operates at (\circnum{1}) or (\circnum{2}) on Triton IR or Triton-GPU
IR is target-independent by construction: it operates on
abstractions that carry no knowledge of the target backends like NVIDIA or AMD, and its output feeds into all the vendor backends
equally. A bit-flip in a data structure that one of these
passes reads or writes is therefore not a bit-flip against a single
vendor's compiled artifact; it is a bit-flip against a representation
that every vendor's backend subsequently consumes. Every pass that
operates at (\circnum{3}) or (\circnum{4}), in contrast, operates on a
vendor-specific representation, and a bit-flip there affects only the
backend whose IR it corrupts.

\begin{figure*}[t!]
    \centering
    \includegraphics[width=\linewidth]{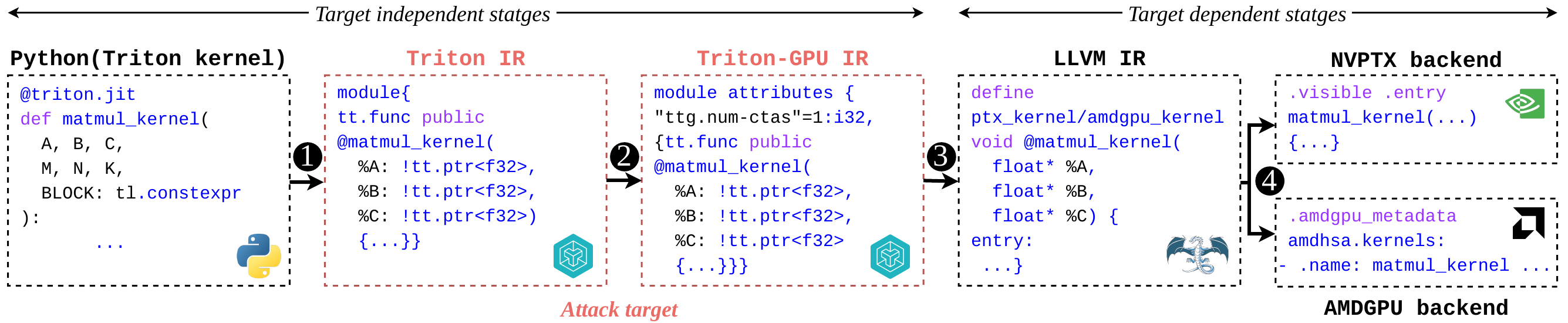}
    \caption{LLVMhammer on the target-independent stages of the Triton compilation pipeline.}
    \label{fig:triton-attack-surface}
\end{figure*}

\subsection{vLLM}

vLLM~\cite{vLLM} is an open-source inference
serving framework designed to run LLMs efficiently. It uses
PagedAttention~\cite{vLLM} to divide the key-value (KV) cache
into fixed-size memory blocks. This design improves GPU memory
utilization, reduces fragmentation, and enables efficient
sharing of the KV cache across concurrent requests, allowing
vLLM to achieve high inference throughput. To execute LLM
inference, vLLM uses Triton to generate optimized GPU
kernels for attention, matrix multiplication, and elementwise
operations such as activation functions and layer
normalization. Triton compiles these kernels through the
MLIR/LLVM compilation pipeline at inference
startup and stores the compiled binaries for later reuse.
Since vLLM depends on Triton-generated kernels, the Triton compilation pipeline
becomes a critical component of every vLLM deployment.

Together, these concepts motivate the attack presented in this
paper: (i) Rowhammer demonstrates that an attacker can induce
bit-flips in DRAM without requiring elevated privileges,
(ii) Triton's compilation pipeline identifies the compiler
intermediate representations that transform high-level GPU
kernels into executable machine code, while distinguishing the
target-independent stages from the vendor-specific backend and (iii) finally, vLLM shows that modern LLM inference frameworks rely
on Triton-generated kernels to execute inference requests.
In sumamry, a Rowhammer attacker can corrupt the
target-independent stages of Triton's compilation pipeline and can generate suboptimal GPU kernels affecting subsequent LLM inference.



\section{LLVMhammer}
\label{sec:attack_overview}
\subsection{Threat model}

\noindent\textbf{Victim.} The victim runs an LLM inference
service on a shared physical host~\cite{yes_one_flip,prowhammer,gpurowhammer}, and uses
Triton as the GPU kernel compiler for GPU inference.
The victim's model weights are inaccessible to co-tenant processes. We consider the victim's model
trustworthy: its weights are free from prior poisoning or
trojan attacks~\cite{yes_one_flip,prowhammer,gpurowhammer}, and the Triton source and MLIR/LLVM
toolchain the victim uses are unmodified and obtained from
trusted providers.

\noindent\textbf{Attacker.} The attacker is a malicious co-tenant
on the same physical host as the victim, consistent with the
cloud co-tenancy threat model established by prior Rowhammer
attacks on ML systems~\cite{deephammer, yes_one_flip,prowhammer,compiled_models_dnn-ndss26}. The
attacker has the following capabilities: (i) it runs an unprivileged user-space CPU process on the same physical
machine as the victim and has no direct read or write access
to the victim's process memory, model weights, or the
Triton/MLIR/LLVM source, (ii) the attacker can allocate
physical DRAM pages and perform Rowhammer to induce bit-flips in physically adjacent DRAM
rows, following the standard Rowhammer attack~\cite{flippingbits,ddr3_ecc_fail,ddr4_ecc_fail,ddr5_ecc_rowhammer,blacksmith,trrespass}, and (iii) the attacker cannot modify the Triton source,
the MLIR/LLVM toolchain, the victim's model weights, or
the victim's inference serving configuration. Triton is
deployed as a shared library, and prior attacks have
established memory deduplication as a mechanism for mapping
identical library pages from different processes to the same
physical DRAM pages~\cite{prowhammer,yes_one_flip,deephammer,deepsteal,compiled_models_dnn-ndss26}.
Thus, the attacker can target a physical page containing
shared Triton code without directly accessing the victim's
process. Even if memory deduplication is disabled, an attacker can use memory massaging to manipulate the victim's memory allocation to map a vulnerable page in the victim's address space, as shown by prior Rowhammer attacks~\cite{prowhammer,anotherflip,rambleed}. 

We mount a Rowhammer-based bit-flip attack on the CPU DRAM to degrade the performance of a GPU-based LLM inference.~\Cref{fig:attack_flow} presents the overview of the LLVMhammer attack against LLM serving applications. The attacker's objective is to increase inference latency metrics such as TTFT and TPOT, and reduce overall throughput measured in tokens and requests per second.

\subsection{Attacking target-independent stages of Triton's compilation pipeline}
 \Cref{fig:triton-attack-surface} shows the target-independent and target-dependent stages of the Triton compilation pipeline. We target the code that generates the target-independent stages (\circnum{1} and \circnum{2}), making the attack vendor-agnostic. At this point, the compilation pipeline remains common across all GPU vendors, and backend-specific optimizations occur only in the later stages (\circnum{3} and \circnum{4}). Consequently, once identified, a vulnerable bit-flip location transfers across Triton deployments without requiring rediscovery and affects kernels targeting both the NVPTX and AMDGPU backends.

 \takeaway{}{Attacking the target-independent stages of the Triton compilation pipeline affects the first target-dependent stage, which is LLVM IR, making our attack GPU vendor agnostic. Thus, we name our attack LLVMhammer.}

\subsection{Finding the useful bit-flips}\label{subsec:useful}

The vast majority of bit-flip locations in \texttt{libtriton.so}, the shared library implementing Triton's compiler functionality, do not produce a useful attack outcome: (i) most flips leave the generated kernel unchanged, (ii) some cause the compiler or generated kernel to crash, and (iii) others corrupt the kernel's output. We therefore define a potentially useful bit-flip as one that simultaneously satisfies the following three properties.


\begin{figure*}[t!]
    \centering
    \includegraphics[width=\linewidth]{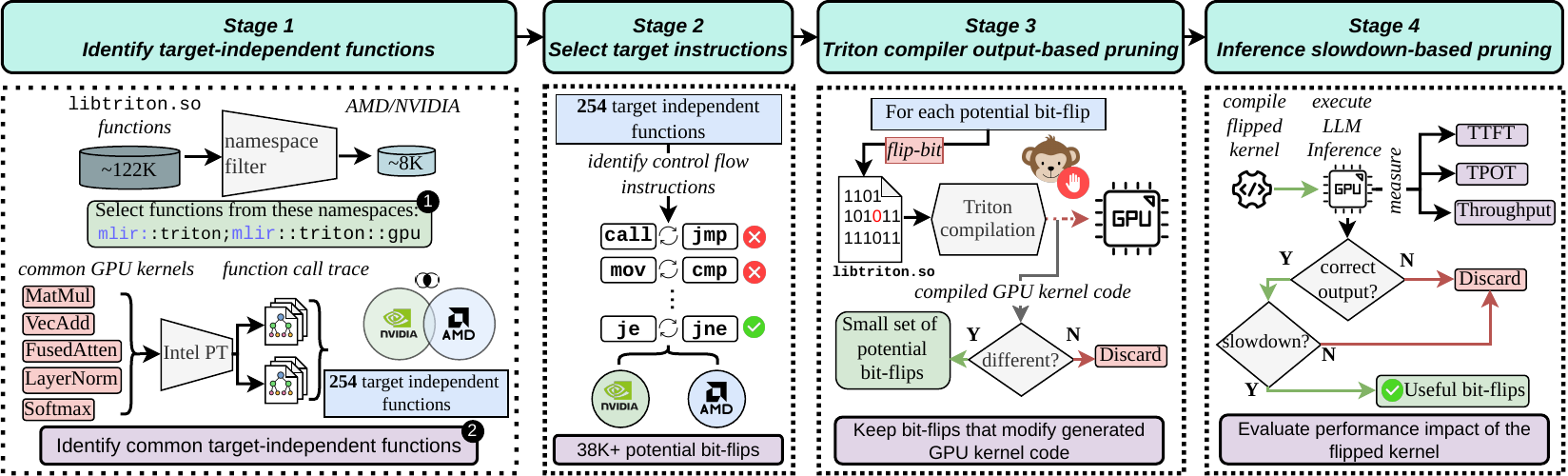}
    \caption{Overview of Karnage’s four-stage pruning of useful bit-flip locations.}
    \label{fig:karnage}
\end{figure*}

\noindent\textbf{P1:Target independence.} The bit-flip must
reside in a function belonging to the target-independent stages
of the Triton pipeline, specifically within the Triton IR or
Triton-GPU IR optimization stages that execute before any
vendor-specific compilation occurs. Because the functions
implementing these stages are shared across all vendor binaries,
a location satisfying P1 generalizes across GPUs from different vendors. A bit-flip that does not
satisfy P1 resides in a vendor-specific function and requires
separate identification for each target backend, forfeiting the
cross-architecture reach that distinguishes this attack.

\noindent\textbf{P2:Performance slowdown without output
corruption or crash.} The bit-flip must cause the Triton compiler to
generate a suboptimal kernel that executes measurably slower than its optimized
counterpart while producing semantically correct outputs and
completing without a crash. A crash is
immediately observable by the victim's inference serving
infrastructure and terminates the inference request, prompting
investigation. Concretely,
the suboptimal kernel must produce a measurable increase in TTFT,
TPOT, or a measurable reduction in tokens per second relative
to the kernel the unmodified compiler produces. 

\noindent\textbf{P3:Consistent slowdown across different inference configurations.}
The slowdown must persist across diverse inference inputs, prompt lengths, and models, consistently increasing both TTFT and TPOT.


 \takeaway{}{A useful flip must lie in a target-independent region of the triton compilation pipeline (P1), cause inference slowdown (P2), and should show consistent slowdown across varying prompts (P3).}

\Cref{subsec:tool} describes the automated methodology the attacker follows to locate useful bit-flip locations in \texttt{libtriton.so} without manual inspection.


\subsection{Karnage: A tool for finding useful bit-flips}
\label{subsec:tool}

\begin{table}[t]
\centering
\caption{Number of functions in
\texttt{libtriton.so} tailored for NVIDIA and AMD GPUs}
\label{tab:functions-libtriton}
\begin{tabular}{lrr}
\toprule
\textbf{Function count} & \textbf{NVIDIA} & \textbf{AMD} \\
\midrule
Total functions in \texttt{libtriton.so}           & 121,735   & 122,148   \\
Target-independent functions (namespace filter)    & 7,963     & 8,219     \\
Target-independent functions (Intel PT filter)     & 254       & 254       \\
\bottomrule
\end{tabular}%
\end{table}

Identifying a useful bit-flip location is the key challenge in mounting LLVMhammer. The attacker must find a useful bit-flip location that satisfies P1, P2, and P3. The Triton compiler library, i.e., \texttt{libtriton.so} contains more than 120,000 functions found using the \texttt{nm} Linux binary utility ~\cite{nm_linux}, and each function contains hundreds of potentially useful bit-flip locations. Evaluating a bit-flip location requires inducing a bit-flip, compiling the kernel, and running a complete LLM inference, which takes approximately 30 seconds. Exhaustively evaluating bit-flip locations would require approximately 12 years, making brute-force search impractical. These constraints require an efficient pruning strategy that quickly eliminates unlikely bit-flip locations while preserving useful ones. We propose Karnage, which implements this strategy by systematically reducing the search space to a small set of high-confidence, potentially useful bit-flip locations. The Karnage pruning strategy is a fully offline pre-step to the LLVMhammer. The attacker then uses the identified bit-flip location to mount LLVMhammer against the target LLM serving application, as shown in~\Cref{fig:attack_flow}. ~\Cref{fig:karnage} summarizes different stages of Karnage. We describe each stage below.

\subsubsection{Stage 1: Identifying target-independent functions}
Karnage first identifies the functions that can contain useful bit-flip locations. LLVMhammer targets LLM-serving applications that compile Triton kernels across different GPU vendors and kernel types. Karnage first identifies the target-independent functions that execute across different kernels and GPU vendors. It then searches these functions for platform-specific bit-flip locations that satisfy P1, P2, and P3. \texttt{libtriton.so} contains more than 120,000 functions, as shown in \Cref{tab:functions-libtriton}, and profiling every function is impractical. 

Karnage identifies target-independent functions in two steps: (i) namespace filtering and (ii) Intel Processor Trace (Intel PT)~\cite{intel_pt_trace} cross-application and cross-platform intersection. In step one, Karnage performs coarse filtering based on function namespaces. Functions in namespaces such as \texttt{mlir::triton} and \texttt{mlir::triton::gpu} implement target-independent optimization passes on Triton IR and Triton-GPU IR. Step one narrows the search space from over 120,000 functions to 7,963 for NVIDIA and 8,219 for AMD, reducing the candidate set by approximately 94\%, as shown in \Cref{tab:functions-libtriton}.

As namespace filtering alone does not guarantee that every selected function executes during the compilation of every Triton kernel. Karnage therefore uses Intel PT to record function-call traces from \texttt{libtriton.so} while compiling representative LLM inference kernels, including matrix multiplication, vector addition, fused attention, layer normalization, and softmax, on both NVIDIA and AMD platforms. Karnage first takes the intersection of the function call traces across all kernels on each platform and retains only the functions that appear in every compilation trace. It then takes the intersection of the surviving functions across both platforms. The resulting 254 functions execute across all evaluated kernels on both GPU vendors and provide a common search space for identifying platform-specific useful bit-flip locations, as shown in \Cref{tab:functions-libtriton}. 

\takeaway{}{Karnage's filtering reduces the search space from more than 120K functions to 254 target-independent functions.
}
 Next, we identify instructions within these functions that can be targeted as bit-flips.

\begin{table}[t]
\centering
\caption{Control-flow altering instruction pairs at
1-bit Hamming distance in the x86-64-based CPU}
\label{tab:instruction-pairs}
\begin{tabular}{llll}
\toprule
\textbf{Instruction. A} & \textbf{Opcode A} &
\textbf{Instruction. B} & \textbf{Opcode B} \\
\midrule
\texttt{je}     & \texttt{0x74} & \texttt{jne}    & \texttt{0x75} \\
\texttt{jl}     & \texttt{0x7C} & \texttt{jge}    & \texttt{0x7D} \\
\texttt{jle}    & \texttt{0x7E} & \texttt{jg}     & \texttt{0x7F} \\
\texttt{jb}     & \texttt{0x72} & \texttt{jae}    & \texttt{0x73} \\
\texttt{jbe}    & \texttt{0x76} & \texttt{ja}     & \texttt{0x77} \\
\texttt{js}     & \texttt{0x78} & \texttt{jns}    & \texttt{0x79} \\
\texttt{jp}     & \texttt{0x7A} & \texttt{jnp}    & \texttt{0x7B} \\
\texttt{cmove}  & \texttt{0x44} & \texttt{cmovne} & \texttt{0x45} \\
\texttt{cmovl}  & \texttt{0x4C} & \texttt{cmovge} & \texttt{0x4D} \\
\texttt{cmovle} & \texttt{0x4E} & \texttt{cmovg}  & \texttt{0x4F} \\
\texttt{cmovb}  & \texttt{0x42} & \texttt{cmovae} & \texttt{0x43} \\
\texttt{cmovbe} & \texttt{0x46} & \texttt{cmova}  & \texttt{0x47} \\
\bottomrule
\end{tabular}
\end{table}

\subsubsection{Stage 2: Selecting target CPU instructions}

Having identified 254 target-independent functions, Karnage
must determine which CPU instructions within those functions are
worth flipping as the entire compilation pipeline gets executed on a CPU. 
Flipping an arbitrary bit in an arbitrary
instruction is unlikely to alter the compiler's optimization
behavior in a meaningful way: a bit-flip in an arithmetic
instruction typically corrupts a single intermediate value
without changing the control flow of the pass, and the effect
either gets discarded by a subsequent operation or produces a
crash rather than a suboptimal but correct kernel.

Karnage instead restricts its attention to
\textit{control-flow altering instructions}~\cite{yes_one_flip},
specifically conditional jump and conditional move instructions
that determine which branch of an optimization pass the
compiler executes. A bit-flip in one of these instructions
changes the branch the compiler takes, causing an optimization
pass to follow a different code path and emit a different
compiled kernel, which is precisely the mechanism by which a
useful flip produces a suboptimal but correct kernel rather
than a crash or an unchanged output.

Critically, many pairs of control-flow altering instructions
lie at a Hamming distance of exactly one in their x86-64 binary
encodings, meaning a single bit-flip in the opcode of one
instruction produces another syntactically valid instruction. \Cref{tab:instruction-pairs} lists all
such pairs that Karnage targets. Restricting the search to
these pairs ensures that every potentially useful bit-flip produces a
valid, executable instruction rather than an illegal opcode
that crashes the Triton process before any kernel compilation
completes. This restriction is what makes the property P2 achievable in
practice: a bit-flip in \texttt{libtriton.so} produces a valid, executable instruction that does not crash the Triton compiler, so the
compiler runs to completion and emits a kernel that can be
evaluated for performance and correctness.

Selecting target instructions from \Cref{tab:instruction-pairs}
to the 254 target-independent functions identified in stage one
yields 38,769 potentially useful bit-flips on NVIDIA and 42,996 on AMD,
as shown in \Cref{tab:functions-libtriton}. As reported in
\Cref{tab:time-reduction}, evaluating all of these potentially useful bit-flips
with a full GPU inference run at 30 seconds per trial would
require approximately 320+ hours per platform. While this
is a substantial reduction from the 9,800+ hours required by exhaustive evaluation, as shown in \Cref{tab:time-reduction}. The search space still benefits from
further pruning before GPU profiling. The next stage eliminates the
GPU execution cost for the majority of the remaining potentially useful bit-flips
by exploiting Triton's offline compilation.

\begin{table}[t]
\caption{Search space reduction and estimated profiling time with subsequent Karnage stages}
\label{tab:time-reduction}
\centering
\resizebox{\columnwidth}{!}{%
\begin{tabular}{l rr rr rr}
\toprule
& \multicolumn{2}{c}{\makecell{\textbf{Potentially useful}\\
\textbf{bit-flip locations}}}
& \multicolumn{2}{c}{\makecell{\textbf{Time per}\\
\textbf{trial (s)}}}
& \multicolumn{2}{c}{\makecell{\textbf{Total}\\
\textbf{time (hrs)}}} \\
\cmidrule(lr){2-3} \cmidrule(lr){4-5} \cmidrule(lr){6-7}
\textbf{Stage}
& \textbf{NVIDIA} & \textbf{AMD}
& \textbf{NVIDIA} & \textbf{AMD}
& \textbf{NVIDIA} & \textbf{AMD} \\
\midrule
Stage 1: Namespace filter
& 1,183,008 & 1,326,122
& 30 & 30
& 9,858 & 11,051 \\
Stage 1: Intel PT filter
& 38,769 & 42,996
& 30 & 30
& 323 & 358 \\
Stage 4: GPU profiling
& 1,176 & 810
& 30 & 30
& 10 & 7 \\
\bottomrule
\end{tabular}%
}
\end{table}

\begin{table*}[!t]
\caption{Effect of a bit-flip on the generated LLVM IR, PTX, and SASS}
\label{tab:flip-codegen}
\centering
\setlength{\tabcolsep}{8pt}
\resizebox{\textwidth}{!}{%
\begin{tabular}{@{} l c c @{}}
\toprule
\textbf{IR level}
  & \textbf{Baseline (unmodified)}
  & \textbf{Flipped} \\
\addlinespace[2pt]
\multicolumn{3}{@{}c@{}}{%
  Vectorized vs.\ Scalar inline-asm emission} \\[4pt]
LLVM IR & \usebox{\llvmbase} & \usebox{\llvmflip} \\
\addlinespace[2pt]
\addlinespace[2pt]
\multicolumn{3}{@{}c@{}}{%
  128-bit coalesced load vs.\ scalar 32-bit loads} \\[4pt]
PTX     & \usebox{\ptxbase}  & \usebox{\ptxflip}  \\
\addlinespace[6pt]
\addlinespace[2pt]
\multicolumn{3}{@{}c@{}}{%
  128-bit LDG/STG vs.\ 16-bit scalar LDG} \\[4pt]
SASS    & \usebox{\sassbase} & \usebox{\sassflip} \\
\addlinespace[6pt]
\bottomrule
\addlinespace[4pt]
\end{tabular}%
}
\footnotesize
\raggedright
\textit{Note:} We have shown the effect of a bit-flip on \emph{LLVM IR, PTX,} and \emph{SASS}, which is specific to NVIDIA, but we observe similar results for AMD.
\end{table*}

\subsubsection{Stage 3: Triton compiler output-based pruning}

After stage two, the surviving potentially useful bit-flips still require
empirical verification to determine which flip locations, when targeted in the CPU's DRAM, actually alter the compiled kernel output. Running a full GPU inference
trial for each of the 38,769-42,996 remaining potentially useful bit-flips at
30 seconds per trial remains infeasible.

Karnage overcomes this bottleneck by exploiting Triton's
\emph{offline compilation} feature~\cite{triton_github}. Rather than triggering a full
inference run for each potentially useful bit-flip, Karnage induces the bit-flip in the Triton library, 
triggers kernel compilation by running the target application,
and intercepts the kernel launch call before any GPU execution
occurs. This halt
is necessary because Triton's optimization passes consume
runtime parameters such as tensor shapes and strides to make
kernel optimization decisions, and these parameters must be
supplied even in offline compilation to obtain a meaningful
compiled code. The monkey-patch (modifies code dynamically at runtime) intercepts the kernel
launch call, supplies the required runtime parameters to the
compiler, and stops execution before any GPU compute occurs, making the entire filtering step CPU-only. This enables parallel execution, significantly reducing the time required for profiling.

Karnage then compares the compiled kernel code produced
under the potentially useful bit-flip against the code produced
by the unmodified compiler. Bit-flips that produce no
changes in the compiled code are discarded, because a flip that
does not alter the compiled code cannot affect inference
performance or output correctness, regardless of how the
kernel executes on the GPU. Only bit-flips that produce an altered compiled code proceed to
stage four, eliminating the majority of the remaining
bit-flips without any GPU execution.

\subsubsection{Stage 4: Inference slowdown-based pruning}

The potentially useful bit-flips that survive stage three alter the
compiled kernel code but have not yet been evaluated for
whether the alteration satisfies P2 and P3 that is, whether
the altered kernel executes slower than the unmodified kernel
without corrupting the inference output or crashing. Karnage
evaluates each surviving potentially useful bit-flip by inducing the bit-flip,
compiling the kernel, and running a full LLM inference
on the GPU, measuring TTFT, TPOT, and tokens per
second under the corrupted kernel and comparing them against
the baseline measurements from the unmodified kernel.
Bit-flips where the corrupted kernel crashes or produces
outputs that diverge from the ground truth are discarded
since they violate P2. Bit-flips where the performance
metrics do not change measurably across the range of
inference inputs tested are discarded since they violate P3.
Bit-flips where the corrupted kernel produces a consistent
and measurable degradation in TTFT, TPOT, ITL, and throughput across the full range of inference inputs without
output corruption or crash are retained as useful flips for
the inference slowdown attack.

\noindent\textbf{What causes the inference slowdown?}
\Cref{tab:flip-codegen} shows the effect of a useful bit-flip
on the generated LLVM IR, PTX, and Streaming Assembler (SASS) for an attention kernel. At the LLVM IR level, the bit-flip causes the
compiler to emit four separate scalar inline-asm calls in place
of a single vectorized call, replacing one \texttt{ld.global.v4.b32}
instruction that loads four 32-bit values in a single transaction
with four individual \texttt{ld.global.b32} instructions that each
load one value. This change propagates directly into the PTX: the
baseline emits 128-bit coalesced loads and stores using
\texttt{ld.global.v4.b32} and \texttt{st.global.v4.b32}, while
the flipped kernel emits one scalar 32-bit instruction per value,
producing four times as many memory transactions for the same
amount of data. At the SASS level the consequence is most stark:
the six 128-bit vectorized global memory transactions
(\texttt{LDG.E.128} and \texttt{STG.E.128}) in the baseline are
absent in the flipped kernel, replaced by 48 half-width
16-bit scalar loads (\texttt{LDG.E.U16}) that did not exist in
the baseline.

The above transformations in LLVM IR, PTX and SASS directly explains the observed slowdown in inference. Modern GPU memory systems achieve peak
bandwidth by issuing wide, coalesced memory transactions that
serve all threads in a warp from a single cache line fetch. A
128-bit \texttt{LDG.E.128} transaction moves 16 bytes per thread
in one memory request; replacing it with eight 16-bit
\texttt{LDG.E.U16} transactions requires eight separate memory
requests to move the same data, each of which must independently
traverse the memory hierarchy and compete for memory bandwidth. The result is a measurable
increase in TTFT, TPOT, and ITL and a corresponding reduction
in throughput, without any change to the arithmetic operations
the kernel performs or the values it produces, which is why the 
flipped kernel remains functionally correct while executing
measurably slower than its optimized counterpart.

 \takeaway{}{A single bit-flip in the Triton library can lead to making suboptimal decisions during code optimization, such as choosing multiple scalar loads over a vectorized load.}

\begin{table}[]
\centering
\caption{Platforms for evaluation}
\label{tab:platform}
\resizebox{\columnwidth}{!}{%
\begin{tabular}{|l|ll|}
\hline
\textbf{Component}                                            & \multicolumn{1}{l|}{\textbf{Platform A}}                                                               & \textbf{Platform B}                                                                 \\ \hline
\textbf{CPU}                                                  & \multicolumn{1}{l|}{Intel i7-4790 (Haswell)}                                                           & Intel i7-8700 (Coffee Lake)                                                         \\ \hline
\textbf{DRAM}                                                 & \multicolumn{1}{l|}{8 GB Kingston DDR3 @ 1600}                                                         & 8 GB Corsair DDR4 @ 2400                                                            \\ \hline
\textbf{OS}                                                   & \multicolumn{2}{c|}{Ubuntu 20.04.6 LTS}                                                                                                                                                      \\ \hline
\textbf{Kernel}                                               & \multicolumn{1}{l|}{5.15.0-131}                                                                        & 6.2.0-060200                                                                        \\ \hline
\multirow{2}{*}{\textbf{GPU}}                                 & \multicolumn{2}{c|}{\multirow{2}{*}{\begin{tabular}[c]{@{}c@{}}NVIDIA RTX 4090 (24 GB GDDR6X)\\ AMD RX 7900 XTX (24 GB GDDR6)\end{tabular}}}                                                 \\
                                                              & \multicolumn{2}{c|}{}                                                                                                                                                                        \\ \hline
\textbf{\begin{tabular}[c]{@{}l@{}}LLM\\ Models\end{tabular}} & \multicolumn{2}{c|}{\begin{tabular}[c]{@{}c@{}}Qwen2.5-7B-Instruct~\cite{qwen}, Llama-3.1-8B-Instruct~\cite{llama}\\ Mistral-7B-Instruct-v0.3~\cite{mistral}, Deepseek-llm-7B-chat~\cite{deepseek}\end{tabular}} \\ \hline
\end{tabular}%
}

\vspace{2pt}

\parbox{\columnwidth}{%
\footnotesize
\textit{Note:} NVIDIA GPU uses CUDA~13.2~\cite{cuda_toolkit}
and vLLM 0.24.0 with Triton 3.6, while AMD GPU uses
ROCm~7.14~\cite{amd_rocm} and vLLM 0.25.1+rocm723
with Triton 3.6.
}

\end{table}


\section{Evaluation}
\label{sec:eval}
In this section, we evaluate the end-to-end LLVMhammer attack using the useful flips identified by Karnage. We measure the impact of LLVMhammer on LLM inference across four LLM models and two GPU vendors. For each model-GPU vendor combination, we perform 30 repeated runs and apply Welch’s t-test~\cite{welch_t_test} to determine whether the degradation in throughput, TTFT, TPOT, and ITL is caused by the bit-flip rather than the inherent variance of batched inference~\cite{batched_inference_noise_article,batched_inference_noise_vllm_issue}.

\subsection{Evaluation Methodology}
\label{sec:eval-methodology}

~\Cref{tab:platform} presents the evaluation platforms for our attack. We perform our experiments on two different platforms and with two different GPU vendors. Although we demonstrate LLVMhammer on DDR3 and DDR4 DRAMs, LLVMhammer would also work on any DDR5 DRAMs, which are susceptible to Rowhammer as shown by prior works~\cite{zenhammer}~\cite{ddr5_ecc_rowhammer}.
 

We evaluate the inference slowdown using the command
\texttt{vllm bench serve}~\cite{vllm_bench_serve}, a standard benchmarking tool
that drives a running vLLM inference server with a configurable
number of prompts and records per request and aggregate
performance metrics. All prompts are drawn from the
ShareGPT~\cite{sharegpt} dataset, which provides realistic conversational prompt lengths representative of production LLM
serving workloads. 
We vary the number of prompts submitted to the server from small to large:
$n \in \{10, 100, 1000\}$~\cite{vAttention,vllm_number_of_prompts}. For each value of $n$, we measure
five metrics: TTFT, TPOT, ITL, and throughput in the form of tokens/second and requests/second. We evaluate all
configurations across four LLM models listed in
\Cref{tab:platform} on both NVIDIA RTX 4090 and AMD RX 7900 XTX.

\begin{table}[t!]
\centering
\caption{Number of metrics (out of five: TTFT, TPOT, ITL,
throughput\textsubscript{req}, throughput\textsubscript{tok})
with $p < 0.05$ under Welch's t-test (baseline vs.\ flipped,
30 runs each). \checkmark~denotes all five metrics significant}
\label{tab:indep-ttest-summary}
\small
\begin{tabular}{ll ccc}
\toprule
\textbf{Vendor} & \textbf{Model}
& \textbf{$n=10$} & \textbf{$n=100$} & \textbf{$n=1000$} \\
\midrule
\multirow{4}{*}{NVIDIA}
& DeepSeek & 2/5 & 3/5 & \checkmark \\
& Llama    & 2/5 & \checkmark & \checkmark \\
& Mistral  & 0/5 & 3/5 & \checkmark \\
& Qwen     & 1/5 & 1/5 & \checkmark \\
\midrule
\multirow{4}{*}{AMD}
& DeepSeek & \checkmark & \checkmark & \checkmark \\
& Llama    & 1/5 & \checkmark & \checkmark \\
& Mistral  & 2/5 & \checkmark & \checkmark \\
& Qwen     & \checkmark & \checkmark & \checkmark \\
\bottomrule
\end{tabular}\end{table}

\begin{figure*}[t!]
    \centering
    \includegraphics[width=\linewidth]{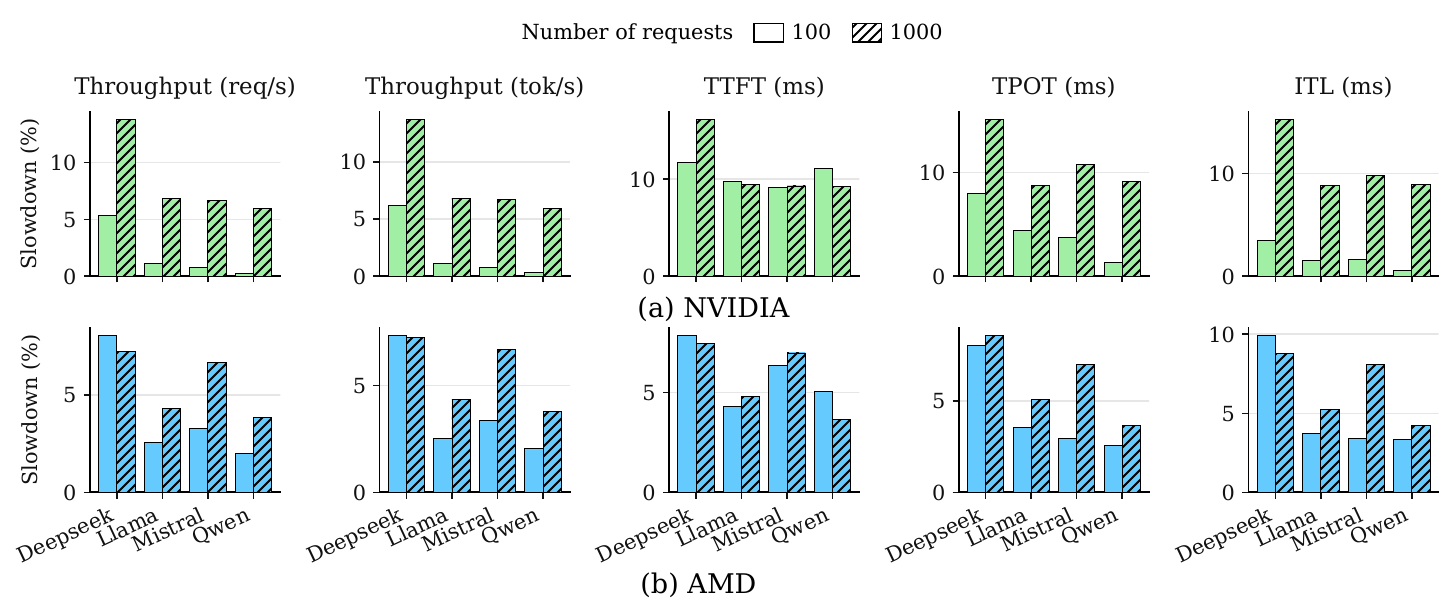}
    \caption{Batched inference slowdown in terms of throughput(requests/second), throughput(tokens/second), TTFT, TPOT, and ITL with varying number of requests for (a) NVIDIA and (b) AMD GPUs.}
    \label{fig:vllm-degration-metrics}
\end{figure*}

\noindent\textbf{Challenges in batched vLLM inference.} 
vLLM manages batching internally through its continuous batching scheduler, which dynamically determines the effective batch size at each forward pass iteration based on KV cache occupancy and request arrival rate~\cite{vllm_continuous_issue,vllm_continuous_article}. 
vLLM's continuous batching scheduler introduces non-trivial
run-to-run variance in all five metrics, even at sampling
temperature $\tau$=$0$~\cite{vllm_continuous_article}, since the effective per-iteration batch
composition varies across runs depending on request timing, KV
cache state, and scheduler decisions. A naive single-run
comparison between baseline and a bit-flipped configuration is
therefore insufficient to establish that any observed
performance slowdown is caused by the bit-flip rather than by this
inherent scheduling noise.

We address this by running each
configuration 30 times under the unmodified Triton binary to
establish a baseline distribution and 30 times under the
bit-flipped Triton binary to establish the performance slowdown.
We then apply Welch's t-test to each metric pair independently
and report the fraction of metrics for which $p < 0.05$, meaning the results are statistically significant~\cite{statistical_significance} and not vLLM's batched inference noise, as
summarized in \Cref{tab:indep-ttest-summary}. Welch's t-test~\cite{welch_t_test}
is appropriate here because it does not assume equal variance
between the two distributions, which matters since the flipped
kernel alters the memory access pattern of the compiled code
and can therefore change the variance of the latency
distribution in addition to its mean.

\noindent\textbf{Statistical validation of inference slowdown} 
\Cref{tab:indep-ttest-summary} shows that at $n$$=$$1000$, every
model on both NVIDIA and AMD produces $p < 0.05$ across all
five metrics, providing conclusive statistical evidence that
the observed slowdown is caused by the bit-flip and not by
scheduling noise. We  evaluated $n$ up to 1000 because production LLM
servers can handle request volumes of this order~\cite{vllm_number_of_prompts},
making this setting representative of realistic inference
workloads. At lower request volumes, the signal becomes weaker and more dependent on the model: at $n$$=$$10$, some models show no
significant slowdown on any metric, while at $n$$=$$100$
significance is partial. This is consistent with the
expectation that a larger request volume sustains higher GPU
utilization for longer, amplifying the cost of the suboptimal
memory transactions the suboptimal kernel emits and producing a
stronger and more consistent slowdown signal relative to the
scheduling noise floor.


\begin{figure*}[t!]
    \centering
    \includegraphics[width=\linewidth]{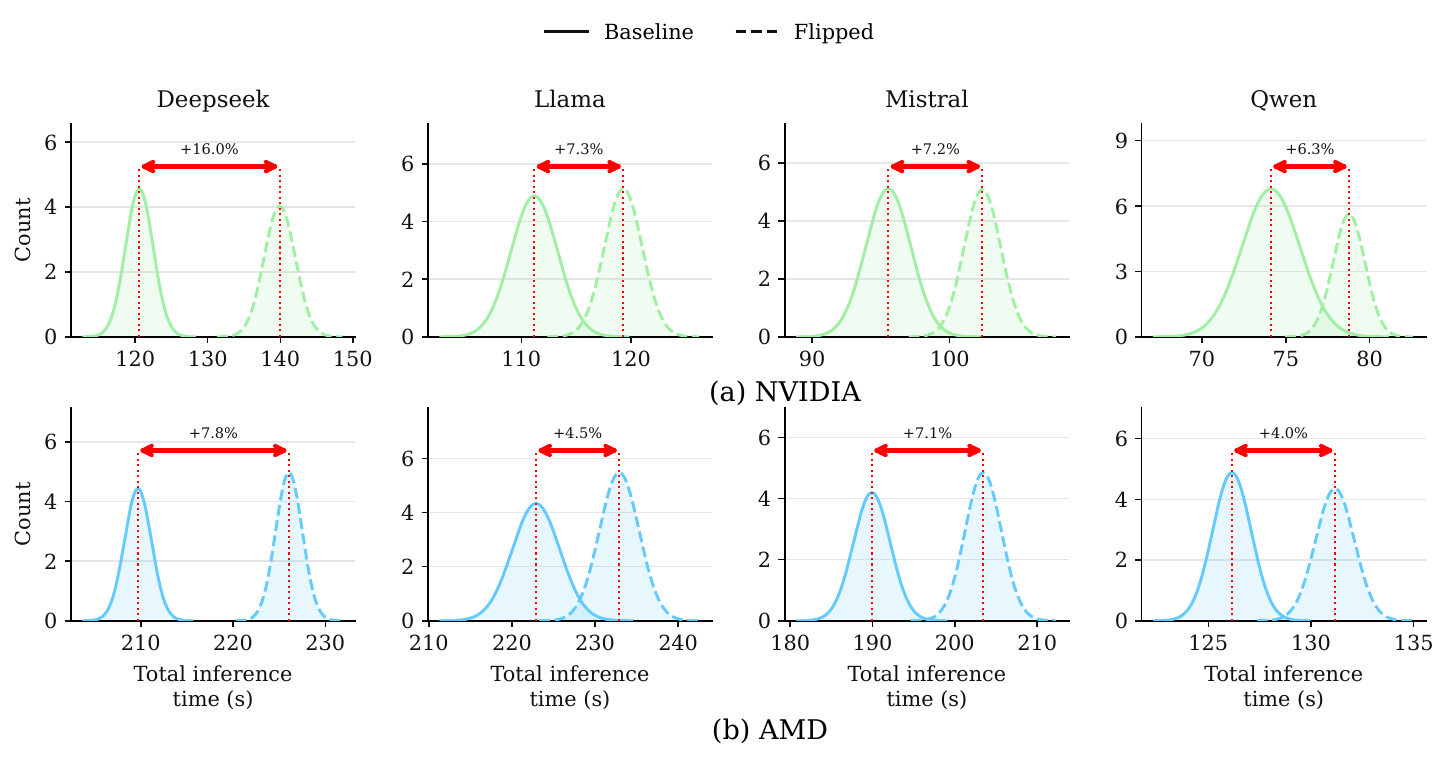}
    \caption{Distribution of total inference time for n=1000 number of requests across 30 runs for (a) NVIDIA and (b) AMD GPUs. Although an increase of 4-16\% does seems modest, in production inference servers which serve millions of requests~\cite{production_llm_load_source,vllm_production_llm_load_source} slowdown becomes quite apparent.}
    \label{fig:vllm-degration-duration}
\end{figure*}

\subsection{LLM Inference Slowdown}

\Cref{fig:vllm-degration-metrics} presents the percentage
slowdown across all five metrics for all four models on
NVIDIA (\Cref{fig:vllm-degration-metrics} (a)) and AMD (\Cref{fig:vllm-degration-metrics}(b)) at $n \in \{100, 1000\}$.
\Cref{fig:vllm-degration-duration} presents the distribution of total
inference time in seconds as a function of prompt count,
comparing the baseline and flipped kernels on both platforms.

\noindent\textbf{Effect of prompt count on inference
slowdown.} As shown in \Cref{tab:indep-ttest-summary} at $n$$=$$10$, the slowdown signal is
inconsistent and unreliable across models and vendors. On
NVIDIA, Mistral shows no significant metric at all, and no
model achieves significance on more than two out of five
metrics. On AMD GPU, the results are more consistent but still
unreliable for Llama and Mistral. The inconsistency arises
because, at low request counts, the inference workload
completes almost instantly. Thus, we cannot reliably distinguish the effect of the suboptimal kernel from vLLM's batched inference noise~\cite{vllm_continuous_article,vllm_continuous_issue}. We therefore exclude $n$$=$$10$
from \Cref{fig:vllm-degration-metrics} and
\Cref{fig:vllm-degration-duration} and focus on
$n \in \{100, 1000\}$.

At $n$$=$$100$, we start seeing more consistent results. AMD shows
consistent significance across all five metrics for three
out of four models, while on NVIDIA, the significance is partial,
with DeepSeek and Mistral reaching 3/5 metrics and only
Llama achieving full significance. The stronger signal on
AMD at $n$$=$$100$ suggests that the memory access penalty
of the flipped kernel manifests more consistently on the
AMD backend at moderate load,

At $n$$=$$1000$, the slowdown is conclusive and consistent
across all models and both vendors. As shown in
\Cref{fig:vllm-degration-duration}, the flipped kernel
increases total inference time by 6.3--16\% on NVIDIA, with
DeepSeek showing the largest slowdown at 16\% and Qwen the
smallest at 6.3\%. On AMD, the increase ranges from 4--8\%,
with DeepSeek again showing the largest effect at 7.8\%. The
gap between the baseline and flipped curves widens decisively at $n$$=$$1000$,
confirming that the slowdown compounds as the number of
requests grows and is unambiguously attributable to the
bit-flip rather than to scheduling noise. Although 4--16\%
may appear modest, production LLM inference servers that
serve millions of requests~\cite{production_llm_load_source,
vllm_production_llm_load_source} accumulate this overhead
continuously, translating a seemingly small per-request
penalty into a high and sustained operational cost.

\noindent\textbf{Impact of inference slowdown at production scale.}
A slowdown of 4-16\% may appear modest in isolation,
but its consequences compound rapidly at the scale of
production LLM inference deployments. A serving
infrastructure that handles one million requests~\cite{production_llm_load_source,vllm_production_llm_load_source} per day
and experiences a 10\% throughput reduction effectively
loses the equivalent of 2.4 hours of serving capacity
every day, or roughly 876 hours per year, without any
output anomaly that would prompt the operator to
investigate. In production deployments where compute
time is billed directly to the customer, this throughput
loss translates to a direct and sustained monetary cost
that accumulates indefinitely as long as the flipped
kernel artifact remains in the Triton on-disk cache. At present, with H100 pricing of \$4.50-\$10.00/GPU-hour~\cite{hf-pricing}, the total cost translates to \$3.9K-\$8.76K per GPU annually, scaling to \$1M-\$2.24M for a 256-GPU deployment. These costs accumulate silently, making even a modest loss of throughput economically significant.
The attack therefore does not need to produce a dramatic
instantaneous degradation;
a modest but persistent slowdown across every inference
session is sufficient to impose a significant operational
cost on the victim.

\newsavebox{\amdgolden}
\newsavebox{\amdflipped}
\newsavebox{\nvidiagolden}
\newsavebox{\nvidiaflipped}

\begin{lrbox}{\amdgolden}
\begin{minipage}{0.44\textwidth}
\begin{lstlisting}[style=flip,
  backgroundcolor=\color{basebg},
  frame=single,
  framerule=0.4pt,
  rulecolor=\color{green!40!black!30}]
A: Immune-mediated thrombocytopenia
B: Thrombotic thrombocytopenic purpura
C: Drug-induced thrombocytopenia ...
\end{lstlisting}
\end{minipage}
\end{lrbox}

\begin{lrbox}{\amdflipped}
\begin{minipage}{0.44\textwidth}
\begin{lstlisting}[style=flip,
  backgroundcolor=\color{flipbg},
  frame=single,
  framerule=0.4pt,
  rulecolor=\color{red!40!black!30}]
,000000000000000000000000000000000
0000000000000000000000000000000000
n000000000000000000000000000000000
\end{lstlisting}
\end{minipage}
\end{lrbox}

\begin{lrbox}{\nvidiagolden}
\begin{minipage}{0.44\textwidth}
\begin{lstlisting}[style=flip,
  backgroundcolor=\color{basebg},
  frame=single,
  framerule=0.4pt,
  rulecolor=\color{green!40!black!30}]
A) Heparin-induced thrombocytopenia
B) Immune thrombocytopenia purpura
C) Thrombotic thrombocytopenic purpura ...
\end{lstlisting}
\end{minipage}
\end{lrbox}

\begin{lrbox}{\nvidiaflipped}
\begin{minipage}{0.44\textwidth}
\begin{lstlisting}[style=flip,
  backgroundcolor=\color{flipbg},
  frame=single,
  framerule=0.4pt,
  rulecolor=\color{red!40!black!30}]
from theise from the, theise, theise
from the, the from the, the from the,
the from the, the from the from theise

\end{lstlisting}
\end{minipage}
\end{lrbox}

\begin{table*}[!t]
\caption{LLM output corruption after a bit-flip on AMD and
NVIDIA GPUs. The same prompt produces coherent output under the
baseline kernel and incoherent output under the suboptimal kernel
on both platforms, demonstrating cross-architecture output
corruption from a single target-independent bit-flip location.}
\label{tab:output-corruption}
\centering
\setlength{\tabcolsep}{6pt}
\resizebox{\textwidth}{!}{%
\begin{tabular}{@{} l l l @{}}
\toprule
\textbf{Platform}
  & \textbf{Baseline output}
  & \textbf{Flipped output} \\
\midrule
\addlinespace[4pt]
\multicolumn{3}{@{}l@{}}{%
  \small\textit{Prompt: A hospitalized 70-year-old woman
  develops thrombocytopenia on day 7 post orthopedic surgery
  $\ldots$ What is the most likely cause?}} \\[6pt]
AMD
  & \usebox{\amdgolden}
  & \usebox{\amdflipped} \\
\addlinespace[4pt]
\midrule
\addlinespace[8pt]
NVIDIA
  & \usebox{\nvidiagolden}
  & \usebox{\nvidiaflipped} \\
\addlinespace[4pt]
\bottomrule
\end{tabular}%
}
\end{table*}

\section{Moving from inference slowdown to inference corruption}
\label{sec:generality}

\Cref{sec:eval} comprehensively evaluates the
effectiveness of our attack in terms of LLM inference slowdown. Now we show that the same target-independent attack target exposes
an additional capability: \emph{corrupting the output of LLM
inference entirely}. Recent works such as
PRowhammer~\cite{prowhammer} demonstrates that GPU-side
inference can be targeted to corrupt model outputs, but its
attack is tightly coupled to NVIDIA's SASS instruction set and
requires the victim's stack to dispatch computation through
cuBLAS. Our attack is not bound by either constraint as the bit-flip locations we identify reside in the
target-independent stages of the Triton compilation pipeline.
The flip location transfers to every GPU vendor's binary
without re-profiling, and the resulting output corruption
manifests on both NVIDIA and AMD platforms from a single
identified location. We repurpose the \emph{Karnage} tool (\Cref{subsec:tool}) to identify useful bit-flip locations for the output corruption attack, where bit-flips only need to satisfy condition P1 (\Cref{subsec:useful}). In this section we demonstrate this
additional capability and establish its cross-architecture reach.


\subsection{LLM Output Corruption}

Beyond inference slowdown, the target-independent
nature of LLVMhammer also supports a different attack goal:
corrupting the semantic content of the LLM's generated text.
A bit-flip in a correctness-controlling pass in the
target-independent pipeline causes the compiler to emit a kernel
that computes incorrect results, producing incoherent or
semantically wrong outputs on every inference request the bit-flipped kernel serves.

\Cref{tab:output-corruption} demonstrates LLVMhammer's capability to corrupt the LLM's output on
both platforms using the same prompt submitted to all four
evaluated models. Under the baseline kernel, both NVIDIA and AMD
produce coherent, clinically relevant responses that correctly
reason about the presented medical scenario. Under the flipped
kernel, outputs degrade entirely: on AMD, the model produces
a stream of repeated numeric characters with no semantic
content, and on NVIDIA, the model produces a looping repetition
of meaningless phrase fragments. Neither output bears any
relationship to the input prompt or to a plausible language
model response, confirming that the bit-flipped kernel corrupts the
computation at a fundamental level rather than introducing
subtle factual errors.

We also measure BERTScore~\cite{bertscore} for the generated texts, following PRowhammer~\cite{prowhammer}. We compare the BERTScore of the golden and flipped executions for all four models on NVIDIA and AMD GPUs. A single bit-flip consistently reduces BERTScore across all models and both GPU vendors, with the flipped outputs reaching scores as low as 0.25 compared with substantially higher scores for the golden executions. Overall, the consistent degradation shows that a bit-flip can alter the semantic content of the generated text, rather than merely affecting inference performance.

The output corruption on both NVIDIA and AMD stems from
bit-flips in the same target-independent functions, even
though each vendor ships a distinct \texttt{libtriton.so}
binary. These functions reside above the vendor-specific
target-dependent functions and are logically identical across all
vendor binaries: identifying a function as a useful flip
target on one platform immediately reveals the same
function as a target on every other vendor's binary,
without repeating the profiling process. The corrupted
compilation decision then propagates into both the NVPTX
and AMDGPU compilation paths equally, since the flipped
functions operate on IR abstractions that carry no
vendor-specific information. 
\takeaway{}{Our attack differs from PRowhammer~\cite{prowhammer} in its ability to identify bit-flip locations that generalize across GPU vendors, whereas PRowhammer identifies exploitable locations by profiling NVIDIA SASS instructions and requires the full profiling process to be repeated for GPU vendors and their instruction sets}

\section{Potential Countermeasures}
\noindent\textbf{LLVMhammer countermeasures.}
LLVMhammer relies on two properties of the Triton
compilation pipeline: the persistence of the compiled kernel
artifacts in the on-disk cache and the integrity of the
\texttt{libtriton.so} binary in memory. Setting the
\texttt{TRITON\_ALWAYS\_COMPILE}~\cite{triton_github} environment variable forces
Triton to recompile every kernel from scratch at each
inference call, bypassing the cache entirely and preventing
any suboptimal kernel from persisting across sessions.
Reloading \texttt{libtriton.so} from disk before each
inference session ensures that any in-memory bit-flip
induced during a prior session does not carry over into
the next compilation. Together, these two mitigations
break the persistence mechanism that makes LLVMhammer
effective, at the cost of increased kernel compilation
overhead at inference startup.

\noindent\textbf{Rowhammer countermeasures.}
Since the attack primarily targets CPU DRAM, conventional CPU-oriented Rowhammer defenses remain applicable. These include DRAM-level and memory-controller-level techniques~\cite{pro_trr_hw_mitigation,rega_hw_mitigation} that rely on hardware modifications. DDR5's Per-Row Activation Counter (PRAC)~\cite{prac_jedec} offers robust Rowhammer protection but introduces timing-channel vulnerabilities~\cite{prac}; the subsequent Timing-Safe PRAC (TPRAC) design eliminates these channels without compromising effectiveness. Error-Correcting Codes (ECC), at hardware or software layers, reduce bit-flip rates but remain susceptible to single-bit corruption~\cite{ddr3_ecc_fail,ddr4_ecc_fail}.

Targeted software defenses such as SoftTRR~\cite{soft_trr_sw_mitigation} and CATT~\cite{catt_usenix2017} protect specific privileged CPU data structures such as page tables and kernel memory, preventing privilege escalation, but do not prevent bit-flips in unprotected DRAM regions. Guard-row approaches like ZebRAM~\cite{zebRAM_osdi2018} alternate guard and data rows to absorb bit-flips but incur a 50-67\% DRAM capacity overhead, making them impractical. Coarse-grained isolation schemes like Siloz~\cite{siloz_sosp2023} assign entire DRAM subarrays to individual domains but limit the number of supported domains and introduce severe memory stranding. The more recent Citadel~\cite{citadel_micro2025} addresses many of these scalability limitations by supporting thousands of variably-sized domains, yet still does not protect against shared-library-based attacks. Memory Band-Aid~\cite{bandaid-ndss2026} instead throttles per-bank DRAM bandwidth in the memory controller to keep row activations below the Rowhammer threshold, but current hardware lacks per-bank limits, forcing a global throttle that incurs overheads of up to 431\%. MOAT~\cite{moat_asplos2025} further shows that Panopticon~\cite{panopticon_dramsec2021}, which inspired JEDEC's PRAC+ABO framework in DDR5, is vulnerable, indicating that PRAC-based mitigations are not inherently secure. Rowhammer remains an unsolved problem as existing mitigations continue to be circumvented~\cite{ddr5_ecc_rowhammer,zenhammer}.

\section{Related Work}

\noindent \textbf{Accuracy degradation attacks.} A large body of research investigates accuracy degradation attacks on ML models~\cite{tbfa,bfa,terminal_brain_damage,yes_one_flip,deephammer,compiled_models_dnn-ndss26,flip_llm_host-2026,genbfa}. Most attacks target model weights and therefore require multiple bit-flips. Several works demonstrate the practical feasibility of such attacks using Rowhammer~\cite{yes_one_flip,deephammer,terminal_brain_damage}. The works most closely related to ours are~\cite{yes_one_flip,prowhammer}, which uses Rowhammer to corrupt library code targeting ML models. Existing fault-injection defenses for DNNs and LLMs, such as DeepDyve~\cite{deepdyve} and BitFlipScope~\cite{bitflipscope_host-2026}, also fail against~\cite{yes_one_flip,prowhammer,compiled_models_dnn-ndss26} and LLVMhammer because these attacks target library code rather than model parameters. ObfusBFA~\cite{ObfusBFA_host-2026}, on the other hand, also tries to mitigate code-based attacks and thus can mitigate FrameFlip~\cite{yes_one_flip} but not PRowhammer, which targets NVIDIA SASS and is outside its scope. LLVMhammer is also resilient against ObfusBFA because it maintains output correctness and is never flagged as vulnerable. 

\noindent \textbf{Rowhammer attacks.} Rowhammer attacks exploit repeated DRAM accesses to induce bit-flips in neighboring memory rows. Early research demonstrated that Rowhammer enables OS privilege escalation~\cite{blackhat15}. Subsequent work extended these attacks to cross-virtual-machine attacks, such as Flip Feng Shui~\cite{flipfengshui}, kernel-manipulation attacks, such as Go Go Gadget~\cite{gogo_gadget}, and hypervisor attacks, such as HyperHammer~\cite{hyperhammer}. A recent large-scale, fully automated study~\cite{flippyram-ndss2026} across 822 systems has shown that Rowhammer remains pervasive across a broad range of real-world, consumer-grade systems. Researchers have also expanded Rowhammer trigger mechanisms beyond traditional CPU-based exploitation. These mechanisms include integrated GPUs that induce bit-flips in shared CPU DRAM~\cite{glitch}, remote network-based attacks using Network Interface Cards~\cite{throwhammer}, FPGA-based attacks~\cite{fpgahammer}, browser-based attacks~\cite{rowhammerjs,sledgehammer}, and ARM/Android platforms~\cite{drammer}.

GPUHammer~\cite{gpurowhammer}, GDDRHammer~\cite{gddrhammer} and GeForge~\cite{geforge} further demonstrate Rowhammer attacks against GDDR6 memory used by GPUs, while PRowhammer~\cite{prowhammer} shows that CPU-side Rowhammer can corrupt GPU workloads through bit-flips in host DRAM. GPUBreach~\cite{gpubreach} demonstrated a Rowhammer-based GPU privilege-escalation attack via page-table tampering, using the resulting arbitrary GPU read/write primitive to leak cryptographic keys, stealthily corrupt ML model code, and even escalate to root privileges on the host CPU despite IOMMU protection.  Rowhammer vulnerabilities also persist across successive DRAM generations, with attacks demonstrated on DDR3~\cite{flippingbits,blackhat15,flipfengshui,hammertime} and DDR4 even in the presence of the TRR mitigation~\cite{trrespass, blacksmith,half_double,smash_rowhammer,posthammer}, LPDDR4~\cite{blacksmith}, and DDR5~\cite{zenhammer,ddr5_ecc_rowhammer}. 
Experimental results further show that ECC-protected DRAM remains vulnerable to targeted bit-flips despite on-die mitigation mechanisms~\cite{ddr3_ecc_fail,ddr4_ecc_fail,ddr5_ecc_rowhammer,column_disturb_2025}.


\section{Conclusion}
In this work, we investigated whether a Rowhammer-induced bit-flip could reduce LLM inference efficiency without affecting output correctness. We presented LLVMhammer, the first Rowhammer attack that caused LLM inference slowdown by inducing a bit-flip in Triton's target-independent compilation pipeline. A single bit-flip in a performance-critical optimization pass caused Triton to generate a correct but suboptimal GPU kernel. Since Triton cached the generated kernel on disk, the slowdown persisted across inference sessions and process restarts without any output anomaly to prompt investigation. We built Karnage, a tool that identifies exploitable bit-flip locations in Triton's target-independent pipeline without manual inspection, turning a one-off finding into a repeatable methodology. As Triton's target-independent stages are shared across GPU vendors, bit-flip locations identified once transferred across NVIDIA and AMD GPUs. We evaluated LLVMhammer on four LLMs across NVIDIA RTX 4090 and AMD RX 7900 XTX GPUs. LLVMhammer increased inference latency by 4-16\%. We further demonstrated that the same attack target corrupted LLM outputs, reducing BERTScore to as low as 25\%. Collectively, these results established that Triton's compilation pipeline represented a previously unexplored and security-critical attack target, and that bit-flips in this pipeline posed a threat to both inference efficiency and output correctness.




%
\balance
\bibliographystyle{IEEEtran}
\bibliography{references}

\end{document}